# Outgassing of Ordinary Chondritic Material and Some of its Implications for the Chemistry of Asteroids, Planets, and Satellites

By

Laura Schaefer[1,3]

and

Bruce Fegley, Jr.[1-3]

Planetary Chemistry Laboratory, McDonnell Center for the Space Sciences

And Department of Earth and Planetary Sciences

Washington University, One Brookings Drive

Campus Box 1169

St. Louis, MO 63130-4899 USA

laura_s@wustl.edu

bfegley@wustl.edu

**Proposed Running Head:**

Outgassing of Ordinary Chondritic Material

**Corresponding Author:**

Laura Schaefer

Department of Earth and Planetary Sciences

Campus Box 1169

Washington University

One Brookings Dr.

St. Louis, MO 63130-4899

laura_s@wustl.edu

Phone: (314) 935 – 6310

Fax: (314) 935 – 7361

**Abstract**

We used chemical equilibrium calculations to model thermal outgassing of ordinary chondritic material as a function of temperature, pressure, and bulk composition and use our results to discuss outgassing on asteroids and the early Earth. The calculations include ~ 1,000 solids and gases of the elements Al, C, Ca, Cl, Co, Cr, F, Fe, H, K, Mg, Mn, N, Na, Ni, O, P, S, Si, and Ti. The major outgassed volatiles from ordinary chondritic material are $CH_4$, $H_2$, $H_2O$, $N_2$, and $NH_3$ (the latter at conditions where hydrous minerals form). Contrary to widely held assumptions, CO is never the major C-bearing gas during ordinary chondrite metamorphism. The calculated oxygen fugacity (partial pressure) of ordinary chondritic material is close to that of the quartz – fayalite – iron (QFI) buffer. Our results are insensitive to variable total pressure, variable volatile element abundances, and kinetic inhibition of C and N dissolution in Fe metal. Our results predict that Earth's early atmosphere contained $CH_4$, $H_2$, $H_2O$, $N_2$, and $NH_3$; similar to that used in Miller – Urey synthesis of organic compounds.

## 1. Introduction

The terrestrial planets have secondary atmospheres (Aston 1924, Russell and Menzel 1933, Brown 1949) that originated through the outgassing of volatile-bearing material during and/or after planetary accretion (e.g., see Lange and Ahrens 1982, Lewis and Prinn 1984, Abe and Matsui 1985, Prinn and Fegley 1987). The outgassed volatiles, which are compounds of H, C, N, O, F, Cl, and S, formed the Earth's early atmosphere and oceans, and the early atmospheres of Mars and Venus. The nature of the early terrestrial atmosphere is of particular interest because it was in this atmospheric environment that the first life forms originated between 4.5 and 3.8 Ga ago (Oró et al. 1990). Whether Mars possessed an early atmosphere conducive for the origin of life is also a matter of intense interest and speculation. Understanding the nature of Venus' early atmosphere is important for understanding whether Venus once had water and lost it or has been a desert planet throughout its entire history.

Unfortunately, the outgassing of volatiles is a largely unexplored topic that is somewhat like the weather – everyone talks about it, but nobody does anything about it. The only prior study is a Masters thesis by Bukvic (1979). He performed chemical equilibrium calculations for gas – solid equilibria in the upper layers of an Earth-like planet. He modeled the planet's composition using H-chondritic or a mixture of H- and C-chondritic material. From this, he calculated the composition of a volcanically outgassed atmosphere, which he found to be in all cases composed of $CH_4 + H_2$. Bukvic never pursued this work further and it did not attract much attention in the field.

We decided to model volatile outgassing because it seemed a natural extension of our prior work on vaporization of high temperature lavas and volcanic gas geochemistry on Io (Schaefer and Fegley 2004, 2005a, b). The thermal outgassing of volatiles during and after planetary accretion is a question of paramount importance for the origin of planetary atmospheres, thermal metamorphism of meteorites, the chemistry and mineralogy of asteroids, the survival of presolar material in asteroids, satellites, and other bodies, the detection of extrasolar planetary atmospheres, and many other topics.

This paper presents the initial results of our work. We used chemical equilibrium calculations to model the thermal outgassing of ordinary chondritic material, i.e., matter with the average chemical composition of ordinary chondrites. The chondrites are undifferentiated (i.e., unmelted) stony meteorites containing metal + sulfide + silicate. The ordinary (H, L, LL) chondrites constitute about 97% of all chondrites. The chondrites are primitive material from the solar nebula and are generally believed to be the building blocks of the Earth and other rocky asteroids, planets and satellites. Chemical equilibrium calculations predict that ordinary chondritic material was abundant in the inner solar nebula where the terrestrial planets formed (e.g., see Barshay 1981, Lewis and Prinn 1984). Geochemical data show that the Earth has chondritic composition and accreted from a mixture of chondritic materials (e.g., Larimer 1971, Wänke 1981, Hart and Zindler 1986, Kargel and Lewis 1993, Lodders and Fegley 1997). For example, the oxygen isotope mixing model predicts that Earth, Mars, and 4 Vesta incorporated ~21%, ~85%, and ~78% H-chondritic material, respectively (Lodders 2000). Thus it makes sense to start modeling the outgassing of the most abundant group of chondrites – the ordinary chondrites.

Section 2 describes our computational methods and data. Section 3 describes the results of our calculations and describes the sensitivity of our results to variations in key input parameters. Section 4 discusses some of the applications of our results to asteroids and the primitive Earth. Section 5 summarizes the major points of this paper. We described preliminary results of our work in three abstracts (Fegley and Schaefer 2006, Schaefer and Fegley 2005c, 2006).

## 2. Methods

### *2.1 Chemical Equilibrium Calculations*

We used chemical equilibrium calculations to model the thermal outgassing of undifferentiated ordinary chondritic material. The chemical equilibrium calculations were done using a Gibbs energy minimization code of the type described by Van Zeggern and Storey (1970). Our calculations consider 20 major rock-forming elements, minor elements, and volatiles in chondrites (Al, C, Ca, Cl, Co, Cr, F, Fe, H, K, Mg, Mn, N, Na, Ni, O, P, S, Si, and Ti). We used the IVTANTHERMO database (Belov et al. 1999), to which we added a number of minerals found in chondrites. In total, our calculations included 224 condensed phases and 704 gaseous species (see Appendix 1 for a list of condensed phases). Our nominal model uses ideal mineral solid solutions. We also did calculations using pure condensed phases instead of mineral solid solutions and separately considered the effects of C and N dissolution in metal.

### *2.2 Bulk Compositions of Ordinary Chondrites*

We used the average (mean) compositions of H, L and LL chondrites in Table 1 for our nominal models. The H, L, and LL nomenclature refers to the total elemental abundance of iron: high total iron (H, ~49% of all ordinary chondrites), low total iron (L, ~36%), and low total iron and low metal (LL, ~14%).The averages in Table 1 were calculated using compositional data from meteorite *falls*. Falls are meteorites that were observed to fall and were recovered at the time. *Finds* are meteorites, which were found at some later time after they fell on the Earth. There are far fewer falls than finds: only ~4.5% of all ordinary chondrites are observed falls. Falls are generally less altered or contaminated than finds. Meteorite finds are subjected to contamination and terrestrial weathering, which alters their composition, mineralogy, and volatile element abundances. Terrestrial weathering corrodes metal, adds carbon and water, and increases the $Fe^{3+}$ content of falls (Jarosewich 1990). For example, the average hydrogen content of H-chondrite finds in the METBASE database (Koblitz 2005) is over 3 times larger than that of falls. Terrestrial weathering also forms clays, carbonates, sulfates, and rust in meteorite falls.

We calculated the average chondrite compositions in Table 1 from data in the METBASE meteorite database (Koblitz 2005), using only analyses in which abundances are given for all of the major rock-forming elements (Si, Mg, Fe, Al, Ti, Ca, Na, K, Ni, Co, P, S, Cr, Mn, O). The bulk compositions of average H-, L-, and LL-chondrite falls were calculated from a total of 90, 123, and 29 analyses, respectively. Table 1 also gives the abundance ranges for each constituent (e.g., the range of $SiO_2$ abundances in the 90 analyses of H chondrites, and so on).

Our results in Table 1 for the average composition of H, L, and LL chondrites are very similar to those reported elsewhere in the literature (e.g., Urey and Craig 1953, Craig 1964, Mason 1965, Jarosewich 1990). The average values of Urey and Craig (1953) show the biggest discrepancies from our values, but even in this case, the disagreements are relatively minor. The differences arise from several factors including more higher quality analyses in recent years, different criteria for including meteorites in an average calculation, meteorite reclassification between the H, L, LL groups since an earlier average was calculated, analyses of new falls, and reporting analytical results with or without volatiles such as water and carbon. We recommend the papers by Urey and Craig (1953) and Mason (1965) for discussion of ordinary chondrite compositions. As described later, we found that compositional variations have only minor effects on our results and do not alter our major conclusions.

*2.3 Volatile Element Speciation and Abundances in Ordinary Chondrites*

The ordinary chondrites contain a number of volatile-bearing phases with variable occurrence and abundance. We computed the average values for H, C, N, F, and Cl in Table 1 using data from the METBASE database. The volatile element abundances in meteorites are variable due to intrinsic variations, analytical problems, contamination, and terrestrial weathering. It can be difficult to determine whether volatiles are indigenous to a meteorite or arise from terrestrial weathering and/or from handling in museums and elsewhere. This is true for both falls and finds (Consolmagno et al. 1998).

Volatiles (H, C, N) are often extracted from meteorites and isotopically analyzed by multi-step heating at increasingly higher temperatures (e.g., Grady and Wright 2003, Hashizume and Sugiura 1997, Yang and Epstein 1983). The fraction of extracted volatiles with Earth-like isotopic compositions is often dismissed as terrestrial contamination. For our considerations here, however, we assume that all volatiles measured for a particular meteorite are indigenous. We do this because there is no guarantee that some of the low temperature “contaminants” are not actually indigenous. This may make our nominal model more volatile-rich than a pristine ordinary chondrite. However, thermal metamorphism on meteorite parent bodies has altered the chondrites now in our meteorite collections. Originally, chondritic material was plausibly even more volatile-rich than even the most volatile-rich ordinary chondrite samples in meteorite collections today. The larger volatile abundances in the unequilibrated ordinary chondrites (i.e., the H3, L3, LL3 chondrites), which are less metamorphically altered than the grade 4 – 6 ordinary chondrites, supports this argument.

To mitigate the effect of taking a volatile-rich nominal model, we have done sensitivity studies on the effects of volatile abundances on our results. We varied the abundances of H, C, O, N, and S over the ranges observed in ordinary chondrites. We describe these results in a later section. We also did calculations using individual analyses of some falls (e.g., Bruderheim L6, Chainpur LL3.4, Homestead L5, New Concord L6) in the METBASE database and noticed no significant difference in our results between the individuals and the average compositions.

*Hydrogen*. Hydrogen occurs as organic compounds and water (absorbed, hydrated salts, hydrous silicates). Kaplan (1971) asserted that there is no indigenous hydrogen in any chondrites except carbonaceous chondrites. However, the subsequent discovery of deuterium-rich material in ordinary chondrites (Robert et al. 1979, McNaughton et al. 1981, Yang and Epstein 1983) shows that at least some of the hydrogen in ordinary chondrites is indigenous. The discovery of fluid inclusions and halite in the Monahans (1998) and Zag H chondrites also shows that indigenous water existed on ordinary chondrite parent bodies (Rubin et al. 2002 and references therein). Table 1 lists $H_2O$ instead of H because most meteorite chemical analyses list $H_2O^-$ (water minus) and $H_2O^+$ (water plus). The $H_2O^-$ is water released by drying at 110 $^oC$ and is loosely bound hydrogen such as physically absorbed water and some of the water in hydrated salts and clays. The $H_2O^+$ is water released by heating at high temperatures and is fixed hydrogen such as H in organic matter, OH in minerals, and more strongly bound water in hydrated minerals (Breger and Chandler 1969).

Our average values for water in H, L, and LL chondrites (see Table 1) agree with those ($H_2O^+$ = 0.32% for H, 0.37% for L, and 0.51% for LL chondrites) from Jarosewich (1990). In contrast, $H_2O$ values from multi-step heating and isotopic analyses of ordinary chondrites range from 0.01 – 0.34% and are somewhat lower than our average values (Robert et al. 1979, 1987a, b, McNaughton et al. 1981, 1982, Yang and Epstein 1983).

*Carbon*. Carbon occurs as graphite, carbides, C dissolved in Fe alloy, organic material, and diamonds. Our average values for the carbon abundances in ordinary chondrites generally agree with those from other sources. Table 1 gives 0.12% (range = 0.01 – 0.34%), Jarosewich (1990) gives 0.11 ± 0.18%, Vdovykin and Moore (1971) give 0.11% (range = 0.02 – 0.35%), and Grady and Wright (2003) give a range of 0.03 – 0.60% for ordinary (H, L, LL) chondrites.

*Nitrogen*. Nitrogen occurs as organic matter, nitrides, N dissolved in Fe alloy, and possibly as ammonium salts (reviewed in Fegley 1983). The small amounts of nitrogen in ordinary chondrites (~ 2 – 120 ppm = μg/g) are difficult to extract due to its inertness (Kung and Clayton 1978). We used nitrogen data from METBASE and other sources (Hashizume and Sugiura 1992, 1995, 1997, Sugiura et al. 1996, 1998, Gibson and Moore 1972, Moore and Gibson 1969, Kothari and Goel 1974). Jarosewich (1990) does not give values for nitrogen. Our average (mean) values (in ppm = μg/g) are 34 (H), 34 (L), and 50 (LL). These are slightly lower than the median values (also in ppm = μg/g) of 48 (H), 43 (L), and 70 (LL) from Mason (1979).

*Fluorine and chlorine*. These elements occur in apatite $Ca_5(PO_4)_3(OH, F, Cl)$ and occasionally as halide salts. Apatite can account for all of the chlorine in chondrites, but only a fraction of the fluorine, of which most must be present in other phases (Mason 1979). The halogens are difficult to analyze (Mason 1971, 1979, Dreibus et al. 1979) and older analyses give higher values than actually present. We used F and Cl analyses from METBASE, Allen and Clark (1977), Dreibus et al. (1979), and Garrison et al. (2000). Mason (1979) gives average values (in ppm = μg/g) of 32 (H), 41 (L), and 63 (LL) for fluorine, slightly higher than our average values in Table 1. Our average values for chlorine are virtually identical to those in Mason (1979).

*2.4 Temperature – Pressure Profiles of Ordinary Chondrite Parent Bodies*

We did most of our calculations using temperature – pressure (T – P) profiles for the asteroid 6 Hebe to model the outgassing of ordinary chondritic material. We also studied the effects of variable total pressure at constant temperature and of variable temperature at constant total pressure on our results. Gaffey and Gilbert (1998) argue that 6 Hebe is a prime candidate for the H-chondrite parent body based on its spectral slope and orbit. Ghosh et al. (2003) developed a thermal model for asteroid 6 Hebe, taking into account both radioactive and accretionary heating. They calculated temperature versus depth profiles for the asteroid at various time intervals since the beginning of accretion (3.7 Ma, 5.7 Ma, and 7.3 Ma after accretion). We used their temperature – depth profiles to calculate the T – P profiles shown in Fig. 1. Using hydrostatic equilibrium, we calculated the lithostatic pressure for the bulk density of Hebe (3.7 ± 1.2 g/cm$^3$) measured by Michalak (2001). We calculated $g$ for a body of the total radius at 3.7 Ma, 5.7 Ma, and 7.3 Ma after accretion. We did not consider the variation of $g$ with depth. For our nominal model, we used the 3.7 Ma T – P profile shown in Fig. 1. All figures hereafter use this T – P profile unless otherwise stated. If we had used the average H chondrite density of 3.2 g/cm$^3$ (Consolmagno and Britt 1998, Consolmagno et al. 1998) we would calculate slightly lower pressures at a given depth. However, as discussed in Section 3.3, slightly lower (or higher) pressures at a given temperature do not significantly change our conclusions.

We used the ideal gas law and average volatile element abundances to see whether using the lithostatic pressure as the total gas pressure is sensible. For example we did a calculation assuming that all carbon in an average H chondrite is in the gas and used $\rho = 3.2$ g/cm$^3$ and porosity = 10% (Consolmagno and Britt 1998, Consolgmano et al. 1998). This gives $P_{gas}$ ~275 bars within the pore spaces at 1225 K, the maximum metamorphic temperature for an H6 chondrite (McSween et al. 1988). If we assume all hydrogen in an average H chondrite is present as $H_2$ gas (as our calculations show), $P_{gas}$ increases to ~1,110 bars. This exceeds the calculated pressure of ~400 bars at the center of 6 Hebe and would break it apart.

Wilson et al. (1999) did a detailed study of internal gas pressures in asteroidal bodies. They found that closed systems typically generate internal gas pressures of tens to thousands of bars, sometimes sufficient to cause explosive disruption of the body. Open systems with extensive fracturing near the surface lose gases to space more efficiently, so the greatest internal gas pressures are found near the center of the body. The calculations of Wilson et al. (1999) were done for carbonaceous chondrites; ordinary chondrites should generate smaller but still substantial pressures. For the sake of this paper, we consider our meteorite parent bodies to be closed systems and assume that

$$P_{gas} = P_{lithostatic} \quad (1)$$

In addition to the above arguments in favor of such large gas pressures, we note that equation (1) is a common assumption in terrestrial metamorphic studies (see for instance Spear 1995). We discuss the effect of total pressure and of open vs. closed systems on our results in sections 3.3 and 3.4.

The maximum temperature we used in our calculations is 1225 K, which is the peak metamorphic temperature for type 6 ordinary chondrites (McSween et al. 1988). This temperature is set by the fact that there is no observed eutectic melting in type 6 ordinary chondrites. Takahashi (1983) melted a Yamato L3 chondrite at pressures from 6 to 30 kbar and found a solidus of 1225 K, defined by the Fe-Ni-S eutectic regardless of pressure for an oxygen fugacity (i.e., partial pressure) equal to the iron – wüstite (IW) buffer. They found the silicate solidus to be ~250 K higher. Since our calculations are set by the thermal model for Hebe ($T < 1225$ K), we are always below the solidus.

*2.5 Comparison to Meteorite heating experiments*

In order to test our calculations, we searched the literature for experiments involving heating of ordinary chondrites and analyses of the gas released. We found remarkably few relevant papers on this subject. Most work on heating and evolved gas analysis was done using carbonaceous chondrites, but these are not the subject of this paper. A number of studies were performed around the turn of the 19$^{th}$/20$^{th}$ century up until the 1920s or so (e.g., Travers 1898-1899, Merrill 1926, Nash and Baxter 1947, Ansdell and Dewar 1886). In principle, these experiments could be useful, but in practice, we cannot use them for comparisons. One reason is that the exact temperature of heating is typically unknown. Samples are described as being heated to a "dull red heat", which may correspond to temperatures between 600 – 900°C. Another reason is that the collected gases are cooled before being analyzed. The cooling probably changes the equilibrium composition of the gas and removes condensable species, such as water vapor, before analysis. Finally, many of the meteorites analyzed are iron meteorites, rather than stony meteorites.

Experiments during the Apollo era looked at gases implanted and trapped in lunar samples (e.g. Gibson and Moore 1972, Gibson and Johnson 1971, Müller 1972, Funkhouser et al. 1971). However, gases trapped inside rocks are different than gases evolved from volatile-bearing phases during heating. Also, the lunar samples are closer to basaltic achondrites than to chondrites.

The most relevant work is that of Gooding and Muenow (1977). They heated a sample of the Holbrook meteorite, an L6 chondrite fall which was recovered in 1912. The heating was done under vacuum in a stepwise fashion; that is, as gas evolved from the chondrite, it was removed and the temperature was gradually increased. Therefore, their experiment is for an open system with evolving bulk composition. We calculated the equilibrium gas chemistry using a bulk chemical analysis of the Holbrook chondrite from Gibson and Bogard (1978). Our results are for a closed system with a constant bulk composition, but are qualitatively similar to those of Gooding and Muenow (1977). They found that $H_2O$ and $CO_2$ were the primary hydrogen and carbon bearing species, with $S_2$ and minor amounts of $SO_2$ becoming abundant above 1073 K. Our results (Fig. 2) show that $H_2O$ and $CO_2$ are the most abundant gases below ~1050 K. From 1050 K to higher temperatures $SO_2$, $H_2O$, and $S_2$ are the three most abundant gases in this order. The equilibrium calculations do not agree with the heating experiment at high temperatures because of the nature of the stepwise heating experiment. By removing the gases released at lower temperatures, primarily $H_2O$ and $CO_2$, the heated Holbrook sample became more reducing and chemical equilibrium shifted from $SO_2$ to $S_2$.

Muenow et al. (1995) heated several ordinary chondrites in a vacuum and measured the gases that were evolved. They found that $H_2O$ and $CH_4$ outgassed below ~900 K. Carbon dioxide formed around 1023 K, and CO was released from ~1100 – 1300 K. Our results indicate that $H_2$ is the most abundant gas, although Muenow et al. (1995) do not mention measuring any $H_2$. We also find that $CH_4$ is only abundant below ~600 K. Above this temperature we find that CO is more abundant than $CO_2$, and $H_2$ is more abundant than $H_2O$. However, due to the experimental set up used by Muenow et al. (1995) and their rapid heating rate, it is unclear that equilibrium is achieved in their system. This is further supported by the fact that their heating experiment did not release all of the volatiles; for instance, their measured abundance of C for the Sharps (H3.4) chondrite was only ~21% of the total carbon abundance as given in the literature (see Table 1 of Muenow et al. 1995).

## 3. Results

### *3.1 Nominal model – gas chemistry*

Figure 3 shows the calculated gas composition for average H-chondritic material along the three temperature/pressure profiles shown in Fig. 1. The temperatures correspond to depth within the parent body, with the coldest temperature at the surface, and the highest temperature at the center of the body. Figure 3a shows the calculated gas composition of average H-chondritic material for the 3.7 Ma Hebe temperature-pressure profile. The major gas at low temperatures is $CH_4$. The abundance of methane begins to drop at ~800 K when graphite becomes a stable solid phase. At ~1050 K, $H_2$ becomes the most abundant gas. Water vapor is always less abundant than $H_2$. Nitrogen is primarily found in $N_2$, with lesser amounts of $NH_3$. The brief peak in $NH_3$ abundance at ~450 K is caused by the decomposition of talc, which releases more hydrogen into the gas phase. However, $H_2$ and $H_2O$ rapidly steal hydrogen away from $NH_3$ as temperature increases. CO and $CO_2$ become more abundant at higher temperatures but are never more abundant than methane. CO is the third most abundant gas at high temperatures and is always more abundant than $CO_2$. All gaseous sulfur, chlorine, and fluorine are found in $H_2S$, HCl, and HF (off scale at lower abundances), respectively.

Figures 4 and 5 are the analogs to Fig. 3a for average L- and LL-chondritic material, respectively. The gas compositions for all ordinary chondritic material are very similar with only minor differences. The small peak in the ammonia abundance observed in H-chondritic material is missing in L- and LL-chondritic material. The associated plateau in the $H_2$ (g) abundance seen in H-chondrites is also missing for L- and LL-chondrites. Both differences are due to the fact that talc is not stable in L- and LL-chondrites but is stable in H-chondrites. The $CH_4/H_2$ crossover point is almost identical for all three types of ordinary chondritic material. Carbon monoxide and $CO_2$ are more abundant (and $CH_4$ less abundant) at high temperatures for both L- and LL-chondrites than for H-chondrites. For LL-chondritic material, CO becomes the second most abundant gas above ~1175 K, and $CH_4$ and $H_2O$ are about equally abundant.

Figure 3b shows the calculated gas composition for H-chondritic material using the 5.7 Ma Hebe temperature-pressure profile. In Fig. 3b, $CH_4$ remains the most abundant gas at all temperatures. The slight decrease in $CH_4$ (and associated increase in $H_2$) above ~900 K is due to the formation of graphite. Talc is present to slightly higher temperatures (~500 K), which causes the broadening of the $NH_3$ peak. Figure 3c shows the gas composition for H-chondritic material gas along the 7.3 Ma Hebe T – P profile. Methane gas is again the most abundant species. Diatomic hydrogen is the second most abundant gas above ~400 K. Water vapor is less abundant than $N_2$ gas at all temperatures. Ammonia is more abundant than $N_2$ gas from 400 – 650 K, at which temperature talc decomposes. Carbon monoxide and $CO_2$ are less abundant than $C_2H_6$.

We do not show the analogous figures for the 5.7 Ma and 7.3 Ma T – P profiles for L- or LL-chondritic material given the close similarities in gas chemistry with H-chondritic material. As before, the major differences between the types of ordinary chondritic material are that the peak in the ammonia abundances and the plateau of the $H_2$ and $H_2O$ abundances seen in H-chondrites are missing from the L- and LL-chondrites.

Figure 6 shows the distribution of the volatile elements (H, C, O, and S) between the major gas and solid species for the calculations shown in Fig. 3a. All nitrogen is in the gas; we discuss its solubility in metal in section 3.7. Figure 6a shows the distribution of hydrogen between the major H-bearing gases ($CH_4$, $C_2H_6$, $H_2$, $H_2O$, $H_2S$, HCl, $NH_3$) and the solid phases talc $Mg_3Si_4O_{10}(OH)_2$ and hydroxyapatite $Ca_5(PO_4)_3OH$. Note the logarithmic abundance scale in Figure 6(a). The major H-bearing species are $CH_4$, $H_2$, and, at low temperatures, talc. Figure 6b shows the distribution of carbon between the major C-bearing gases ($CH_4$, CO, and $CO_2$) and graphite. At equilibrium, $CH_4$ is the major carbon-bearing species at all temperatures. (The fact that any carbon remains in chondrites shows they are disequilibrium assemblages.) We consider the effect of C dissolution in metal in section 3.6. Figure 6c shows the distribution of oxygen between solid phases (silicates and oxides, see section 3.2) and the major gases $H_2O$, CO, and $CO_2$. The amount of oxygen found in the gas phase is insignificant at low temperatures and is less than 0.3% of total oxygen at the highest temperatures.

Figure 6d shows the distribution of sulfur between troilite (FeS) and the gas phase ($H_2S$). The amount of sulfur in the gas is insignificant (<0.1%) at all temperatures. As long as both Fe metal and troilite are present the total amount of sulfur in the gas is regulated by the iron – troilite sulfur fugacity buffer. The equation and $S_2$ fugacity for this buffer are (Fegley and Osborne 2006)

$$\text{FeS (troilite)} = \text{Fe (metal)} + \tfrac{1}{2}\,\text{S}_2\text{ (g)} \qquad (2)$$

$$\log f_{S_2}(bar) = 13.7393 - \frac{17{,}308}{T} - 2.2007\log T + \frac{0.1160(P-1)}{T} \qquad (3)$$

For example, at our highest T – P point along the 3.7 Ma thermal profile (1225 K, 88.8 bars) the sulfur fugacity $fS_2 \sim 6.4 \times 10^{-8}$ bar for reaction (2). The reaction of $H_2$ with this $S_2$ vapor controls the equilibrium partial pressure of $H_2S$.

Figure 7a shows the regions in T/P space in which $CH_4$, CO, and graphite are the most abundant C-bearing compounds within H-chondritic material. The solid lines are equal abundance lines. The dotted line continues the $CH_4$/CO equal abundance line within the graphite-dominant field. The shaded region represents the temperatures and pressures for the three thermal profiles of 6 Hebe (see Fig. 1). As shown in Figure 7a, methane is always the most abundant C-bearing gas along our nominal T – P profile. The tip of the shaded area closest to the $CH_4$/CO boundary is the peak metamorphic temperature (1225 K). It is entirely plausible that, for a slightly different accretion scenario in which the peak metamorphic temperature is reached at shallower depths (lower pressures), 6 Hebe lies at least partially within the CO stability field. However, over time, the asteroid would evolve into the $CH_4$ field (assuming a closed system).

We made similar calculations for average L- and LL-chondritic material. Our results show that 1:1 contours in L- and LL-chondritic material move to slightly lower temperatures (at a given pressure) than for H-chondritic material. Assuming the same T – P profile as for 6 Hebe, the L-chondrite parent body is always within the $CH_4$ field. However, the LL-chondrite parent body lies marginally within the CO field.

Figure 7b shows the equal abundance lines for $N_2/NH_3$ for average H-chondritic material. Ammonia is more abundant than $N_2$ in a small region of P – T space that partially overlaps the P – T profiles used for 6 Hebe. The abundance of $NH_3$ in H-chondritic material seems to be tied to the presence of talc ($Mg_3Si_4O_{10}(OH)_2$), shown with a dotted line. Talc is stable above this line at low temperatures and high pressures. However, this is not true for L- and LL-chondritic material. We found no conditions under which ammonia was more abundant than $N_2$ for L- or LL-chondritic material.

Figure 8 shows the calculated oxygen fugacities ($fO_2$)for average H-, L-, and LL-chondritic material as a function of temperature compared to some common oxygen fugacity buffers and measurements of oxygen fugacities of heated meteorite samples. (The figure caption explains the abbreviations for the different oxygen fugacity buffers.) The oxygen fugacity of average H-chondritic material (Fig. 8a) lies slightly below the quartz-fayalite-iron (QFI) buffer, and the oxygen fugacity of average L-chondritic material (Fig. 8b) lies directly on the QFI buffer. The equation and oxygen fugacity for the QFI buffer are (Fegley and Osborne 2006)

$$Fe_2SiO_4 \text{ (fayalite)} = 2Fe \text{ (metal)} + SiO_2 \text{ (quartz)} + O_2 \text{ (g)} \quad (4)$$

$$\log f_{O_2}(\text{bar}) = 4.5468 - \frac{29{,}194}{T} + 0.8868 \log T + \frac{0.0497(P-1)}{T} \quad (5)$$

The fayalite and metal involved in this buffer in ordinary chondritic material are present in solid solution and are not present as pure phases. Pure quartz is not present but the coexistence of olivine and pyroxene buffers the $SiO_2$ activity at a given temperature. The oxygen fugacity of average LL-chondritic material (Fig. 8c) is slightly larger than that of L-chondritic material and lies slightly above the QFI buffer. It is reasonable to expect that the oxygen fugacity of ordinary chondritic material should be near the QFI buffer.

Brett and Sato (1984) measured the intrinsic oxygen fugacities of two H-, one L-, and two LL-chondrites. The points on the graphs show their results. Our calculated oxygen fugacities agree very well with their results. An abstract by Walter and Doan (1969) reports intrinsic oxygen fugacities of one H- and one L-chondrite. For the H-chondrite, their measurements are significantly lower than those of Brett and Sato (1984) and our calculated fugacities. Their measured fugacities for the L-chondrite Holbrook agree better with, but are still lower than, the data of Brett and Sato (1984). McSween and Labotka (1993) calculated oxygen fugacities for average H4-H6 and L4-L6 chondrites from the compositions of coexisting olivine, orthopyroxene, and metal using

$$2Fe_2SiO_4 \text{ (fayalite)} = 2FeSiO_3 \text{ (ferrosilite)} + 2Fe \text{ (metal)} + O_2 \text{ (g)} \qquad (6)$$

(The lines shown on the graphs are extrapolations of their data to higher temperatures). Calculated oxygen fugacities from McSween and Labotka (1993) are lower than the measured values of Brett and Sato (1984) and our calculated $fO_2$ values. Their results for H chondrites agree at lower temperatures with the oxygen fugacities for one H chondrite measured by Walter and Doan (1969).

*3.2 Nominal model - mineralogy*

The major minerals found in ordinary chondrites are: olivine $(Mg,Fe)_2SiO_4$, Ca-poor pyroxene $(Mg,Fe)SiO_3$, plagioclase $(Na,K,Ca)(Al,Si)Si_2O_8$, diopside $CaMgSi_2O_6$, chromite $FeCr_2O_4$, chlorapatite $Ca_5(PO_4)_3Cl$ and/or whitlockite $Ca_9MgH(PO_4)_7$, kamacite α-(Fe,Ni), taenite γ-(Fe,Ni), and troilite FeS (Rubin 1997). The abundances of Fe-bearing silicates increase from H to L to LL-chondrites whereas the abundance of metal decreases from H to LL. Carbides are rare and occur primarily as cohenite ($Fe_3C$) in unequilibrated ordinary chondrites (Krot et al. 1997). However, the distribution of carbon between carbides, graphite, organic matter, diamond, and other phases in ordinary chondrites is generally unknown.

Our calculated compositions contain all of the above minerals with the exception of whitlockite. We also found talc $Mg_3Si_4O_{10}(OH)_2$, which is a phyllosilicate, and graphite to be stable. Talc typically occurs in meteorites as a terrestrial weathering product. However, phyllosilicates occur in the Bishunpur and Semarkona unequilibrated ordinary chondrites (Alexander et al. 1989, Keller 1998). Brearley (1997) observed talc in the Allende CV3 chondrite, and Rietmeijer and Mackinnon (1985) observed talc in a chondritic interplanetary dust particle. Graphite and poorly graphitized carbon occur in ordinary chondrites (Ramdohr 1973, Grady and Wright 2003, Mostefaoui et al. 2005), but are uncommon. Ramdohr (1973) reports that the graphite in the Grady H3 chondrite "resembles that produced by the thermal decomposition of hydrocarbons in retorts and coke ovens, and must have been produced under comparable conditions."

Figure 9 shows the calculated mineralogy of average H chondritic material for the 3.7 Ma Hebe T – P profile. These calculations included ideal solid solutions for olivine $(Mg,Fe,Mn)_2SiO_4$, orthopyroxene $(Mg,Fe,Ca,Mn)SiO_3$, metal (Fe,Ni,Co), and feldspar $(Na,K,Ca)(Al,Si)_2SiO_8$. We did not model clinopyroxene solid solutions but included pure diopside $CaMgSi_2O_6$, hedenbergite $CaFeSi_2O_6$, and acmite $NaFeSi_2O_6$ in our calculations. Minor phases that are stable but not abundant enough to be visible in Fig. 9 are talc (T < 500 K), graphite (T > 800 K), and sodalite (950 – 1175 K). The Fe-oxides include chromite ($FeCr_2O_4$) and ilmenite ($FeTiO_3$).

Table 2 gives the compositions of the olivine, pyroxene, and feldspar solid solutions. Our calculated olivine compositions for average H- and L-chondritic material are more fayalitic at lower temperatures and more forsteritic at higher temperatures than observed in H and L chondrites. Olivine compositions for average LL-chondritic material match the observed values in LL chondrites. Our calculated pyroxene compositions match observations well at lower temperatures but are too Fe and Ca-rich at higher temperatures. There is no temperature at which we can match both olivine and pyroxene compositions simultaneously for any of the ordinary chondrites. Our feldspar compositions are too Ca-rich at all temperatures, especially at high temperatures. At high temperatures, albite transforms into sodalite $Na_8Al_6Si_6O_{24}Cl_2$ and nepheline $NaAlSiO_4$.

Table 3 compares the normative mineralogies for average ordinary chondrites from McSween et al. (1991) with our calculated mineralogies. We find that there is good overlap between our calculated mineralogies and the normative mineralogies. The amounts of minor phases such as apatite, chromite, and ilmenite are essentially identical, although apatite is slightly higher than the norm in LL-chondrites. For the feldspathic constituents, we have good agreement with the orthoclase abundance at all temperatures. Our albite abundance at lower temperatures agrees with the normative albite abundance; however, as temperature increases, our albite abundance decreases and becomes significantly lower than the normative abundance. Our anorthite abundance does not change significantly with temperature. It is slightly higher than the normative anorthite abundance for H- and LL-chondrites, but agrees well for L-chondrites. Our calculated abundances of orthopyroxene solid solution increase with temperature and match the normative hypersthene value at ~700 K for H- and L-chondrites and ~750 K for LL-chondrites. However, the calculated pyroxene compositions at these temperatures are significantly more Fe-rich than the observed values listed in Table 2. Our olivine abundance decreases with temperature and matches the normative value at ~1100 K for H-chondrites, ~1000 K for L-chondrites, and ~1020 K for LL-chondrites. The olivine compositions at these temperatures fall within the observed range of compositions listed in Table 2. Our calculated metal abundances decrease slightly with temperature and are somewhat lower than the normative metal abundances. Our calculated troilite abundance is approximately constant with temperature for H- and L-chondrites and increases with temperature for LL-chondrites.

If we artificially set both the olivine and orthopyroxene compositions to the average values observed in H-chondrites, we match the normative mineralogy almost exactly. The greatest deviations from the norms is the metal abundance (16.8 – 17.9 mass%), which is again slightly less than the average observed value (Table 3), but still within the range of values observed.

### *3.3 Effect of Variable Total Pressure*

As we stated earlier, we assumed that the total gas pressure is equal to the lithostatic or confining pressure. However, the gas pressure could be greater or less than the lithostatic pressure in other situations. For instance, an impact into a planetesimal's surface could cause fracturing, which would allow gas to escape from the interior, lowering the pressure. Additionally, asteroids should have a small but finite permeability, which allows gas to escape from the interior and reduces the internal gas pressure. We therefore decided to look at the effect of total pressure on the calculated gas chemistry.

Figure 10 shows the calculated gas chemistry for average H chondritic material as a function of pressure at constant temperatures of 600, 900, and 1200 K. Note the logarithmic scale for the total pressure. At all three temperatures, $H_2$ and $CH_4$ are the most abundant gases. $H_2$ is the most abundant gas at low pressures and high temperatures, whereas methane is the most abundant gas at low temperatures and high pressures. As we discussed earlier, the $CH_4/H_2$ crossover point occurs at higher pressures at higher temperatures (see the 900 and 1200 K panels). The abundance of water vapor decreases with increasing pressure and then levels off (see the 900 and 1200 K panels). The $H_2O$ abundance also decreases slightly with increasing temperature (compare the 900 and 1200 K panels). Carbon monoxide and $CO_2$ become abundant gases at high temperatures and low pressures (1200 K panel). The $CH_4/CO$ intersection at 1200 K in Figure 10 corresponds to the position of the 1:1 $CH_4/CO$ contour in Figure 7a. Ammonia and $N_2$ are less abundant at higher temperatures because there are more volatiles in the gas phase. We found similar results for average L- and LL-chondritic material.

Figure 11 shows the calculated oxygen fugacity for average H chondritic material as a function of temperature at pressures ranging from 1 bar to 10 kilobars. As can be seen in the figure, the oxygen fugacity is independent of pressure at temperatures greater than 150 K. This is expected when a solid-state buffer controls the oxygen fugacity because of the relatively small volume change between the different solid phases involved in the buffer (e.g., see Eq. 5 for the QFI buffer).

*3.4 Closed system vs. Open system*

An open system reduces the total internal gas pressure and allows volatiles to escape, which alters the bulk composition of the system. To simulate an open system, we did a series of calculations in which we moved outgassed material from deeper layers upwards, re-equilibrated the volatiles within the next higher layer, and moved the outgassed volatiles upwards again. In this way, we transported essentially all volatiles from the interior of the parent body to its surface. Figure 12 shows the gas composition for our open system calculation. The major gas is $H_2$ at all levels, followed by $CH_4$, $H_2O$, and $NH_3$. The mineralogy at the lower levels is essentially the same as in the closed system. In the layers near the surface, the mineralogy changes and spinel $MgAl_2O_4$, magnetite $Fe_3O_4$, and muscovite $KAl_3Si_3O_{10}(OH)_2$ are stable.

*3.5 Effect of volatile element abundances on gas chemistry*

As we discussed earlier, the volatile contents of meteorites may be significantly affected by terrestrial contamination. In order to understand the effect of volatile content on the gas chemistry, we did calculations for the entire range of volatile contents measured in the ordinary chondrites. We did this by using the average H chondrite abundances given in Table 1 and substituting a different abundance for one of the volatiles (H, C, N, O, and S) at a time. The ranges of volatile abundances used are given in Table 1. All calculations are done using the 3.7 Ma thermal profile in Figure 1. Our results are given in Fig. 13-16 and are discussed below.

*Hydrogen.* Figure 13 illustrates the effect of H abundance on gas chemistry. We found no difference whether we varied hydrogen as pure H or as $H_2O$. The top panel of Fig. 13 shows the gas chemistry for a hydrogen abundance of 2.8 mg/g (2.5% $H_2O$) with all other elemental abundances being those given in Table 1. With this large hydrogen abundance, $H_2$ becomes the most abundant gas at all temperatures and pressures. The second most abundant gases are $CH_4$ below ~900 K and $H_2O$ at higher temperatures. Methane is always more abundant than CO. The second panel is our nominal model (see Figure 3a). In the third panel, the hydrogen abundance is 0.01 mg/g (0.0089% $H_2O$). There is significantly less matter in the gas phase, such that $N_2$ gas becomes the most abundant gas. The second most abundant gases are $CH_4$ and $H_2$. Methane is more abundant at temperatures below ~850 K, and $H_2$ is more abundant at higher temperatures. Water vapor is also fairly abundant and becomes more abundant than $CH_4$ above ~1050 K. Methane is more abundant than CO until ~1200 K, above which CO is more abundant. The bottom panel shows the gas chemistry for an H abundance of 1 μg/g (8.9 μg/g $H_2O$), which is less than the smallest measured abundance of H in ordinary chondrites. We did this computation to determine what happens if Kaplan's assertion that all hydrogen in ordinary chondrites is due to terrestrial contamination is correct. For this computation, most carbon is in graphite. Therefore, the only significantly abundant gas is $N_2$, with minor amounts of CO and $CO_2$ at high temperatures.

*Carbon.* Figure 14 shows the effect of C abundance on gas chemistry. The top panel of Fig. 14 shows the gas chemistry for a carbon abundance of 3.4 mg/g (0.34% C). The gas chemistry is very similar to the average case, shown in the middle panel. The major difference is that, in the higher C case, the $H_2$ abundance does not level out between 500 – 850 K, and there is no peak of $NH_3$ gas, similar to the behavior of volatiles outgassed from average L-chondritic material (Fig.4). In the average case, the sudden rise and plateau of the $H_2$ abundance is caused by the decomposition of talc, but in the high C case, talc is never stable, so we do not see the same effect. The gas abundances at higher temperatures are essentially identical, including the $CH_4/H_2$ cross-over point. The bottom panel of Fig. 14 shows the gas chemistry for a carbon abundance of 0.1 mg/g (0.01% C). Below ~500 K, the most abundant gases are $CH_4$ and $N_2$. Above this temperature, talc decomposes, and $H_2$ becomes the most abundant gas. Methane remains the second most abundant gas until ~750 K, when $H_2O$ becomes more abundant. Nitrogen gas converts into ammonia at ~500 K, and both fall off in abundance at higher temperatures. Carbon monoxide is never significantly abundant.

*Nitrogen.* Table 1 gives a range of nitrogen contents for H chondrites of 2 – 121 ppm (= μg/g). Using this range, we examined the effect of varying nitrogen composition on the bulk gas chemistry (Fig. 15). The middle panel in this plot is the nominal case. Changing the nitrogen abundance only changes the abundances of the major N-bearing gases and has no effect on the other major gases, such as $CH_4$, $H_2$, and $H_2O$. For the higher N abundance, $N_2$ is more abundant, and ammonia does not become more abundant than $N_2$, even at the point of talc decomposition. For the low N abundance, both $N_2$ and $NH_3$ are less abundant, and are off scale the bottom of the graph.

*Oxygen.* Figure 16 shows the variation of gas chemistry with oxygen abundance. Oxygen abundances found in H-chondrites range from 27.7% (Rose City H5 chondrite) – 37.2% (Conquista H4 chondrite). The mean and median oxygen elemental abundances in H chondrites are essentially identical and are 33.55%. Oxygen is typically not determined directly, but rather by difference after all other elements have been measured. Most H-chondrites have oxygen abundances within the range 32 – 36 wt%. Within this range, there is no significant difference in gas chemistry, as can be seen in the second and third panels of Fig. 16. There are only a few meteorites with oxygen abundances outside this range. For H chondrites with larger oxygen abundances, up to 38 wt%, the abundances of $NH_3$, CO, and $CO_2$ increase and the $CH_4/H_2$ crossover point shifts to a slightly higher temperature (~1000 K). At lower oxygen abundances, the gas phase chemistry changes drastically, as can be see in the bottom two panels of Fig. 16. The solid phase chemistry also changes dramatically. For these cases, there is insufficient oxygen available to form silicates. Instead, normally lithophile elements form sulfides (CaS, MgS, CrS and TiS), schreibersite ($Fe_3P$), silicides ($Cr_2Si$, $TiSi_2$), Si metal, SiC, and sinoite ($Si_2N_2O$). These phases are not typically found in ordinary chondrites but are often found in enstatite chondrites and achondrites.

*Sulfur*. We used the range of sulfur abundances (8.2 – 35.2 mg/g) to check if the sulfur abundance had any effect upon the gas chemistry, as we have done above with hydrogen and carbon. The abundances of the stable solid phases changed somewhat, primarily troilite and iron metal. For the low sulfur abundance, troilite was less abundant and iron metal was consequently more abundant, and the reverse for the high sulfur abundance. The major gas phase chemistry is identical for all sulfur abundances.

### *3.6 Solubility of Carbon and Nitrogen in Metal*

As mentioned in section 2.3, carbon and nitrogen are often found dissolved in chondritic metal. We included graphite, $Fe_2C$, $Fe_3C$ (cohenite or cementite), and $Fe_4N$ (roaldite) in our nominal model. We found that $Fe_4N$ was never a stable phase, which indicates to first approximation that nitrogen dissolution in metal should be minor. Nitrogen dissolves in metal as monatomic N and we calculated its solubility using data from Chipman and Elliott (1963) as described in Fegley (1983). We assumed that all nitrogen is found in either $N_2$ (g) or is dissolved in metal, which is a valid assumption at high temperatures. At low temperatures, the solubility of N in metal is negligible. At higher temperatures, we found that the amount of nitrogen dissolved in metal increases, but is still well within the variability of the total nitrogen content of ordinary chondrites.

Figure 17a shows the results of our calculations. Using the average nitrogen abundance of 34 ppm (= μg/g) in H chondrites, we found that ~24% of total nitrogen (~8 ppm) is dissolved in Fe metal at 1225 K. This leaves ~26 ppm nitrogen in the gas. With decreasing temperature, the amount of dissolved nitrogen decreases and becomes insignificant (<1% of total nitrogen) below 900 K. Using a more sophisticated solution model, Hashizume and Sugiura (1998) found 5 – 30 ppm (= μg/g) of N dissolved in taenite (γ-Fe,Ni), regardless of the initial nitrogen concentration, which agrees well with our result. Our sensitivity calculations, where all nitrogen was found in the gas, showed that there was little effect on gas chemistry for nitrogen abundances of 2 – 121 ppm.

Solubility of carbon in metal is more important for our applications because the abundance of carbon in the gas phase is more important for determining the major gas chemistry. It is also consequently a much more difficult problem because carbon occurs in many more compounds than nitrogen. In our nominal model, we found that graphite is stable at high temperatures. We used data on γ-Fe-C alloys from Richardson (1953) to calculate the approximate effect of carbon solubility on our system. (Richardson's data are very similar to those of Chipman (1972) but are in a more convenient form to use). As with nitrogen, we found large amounts of carbon dissolved in Fe at high temperatures, decreasing with temperature. Results are shown in Fig. 17b. At 1225 K, we found ~5000 ppm (0.5 %) C in Fe metal (~65 % of total carbon). This concentration drops to ~500 ppm (0.05 % C) by 800 K, and ~1 ppm ($10^{-4}$ % C) by 600 K. There was no appreciable amount of C in metal at lower temperatures. No iron carbides were stable when carbon dissolved in Fe. Our calculated concentrations are somewhat higher than observed in ordinary chondrites. However, meteoritic metal typically contains Ni, Co, and P. Solubility experiments show that the presence of Ni, Co, P, and other elements decreases the solubility of C and N (e.g., see Wada et al. 1971). We therefore consider our solubility calculations to be an upper limit on the amount of C dissolved in metal. We found that even with such large C concentrations in metal, there was no discernible effect on the calculated gas chemistry, i.e. the major volatiles are still $H_2$ and $CH_4$ despite the large amount of C in metal. This is in stark contrast to the results of Hashizume and Sugiura (1998), who found that CO was the primary C-bearing gas, and $CH_4$ was never abundant.

## 4. Discussion

We briefly discuss some of the applications of our results to outgassing on asteroids and the early Earth below.

### *4.1 Asteroidal Out-gassing and Metamorphism of Ordinary Chondrites*

The concept of meteorite metamorphism dates back at least as far as Merrill (1921). Dodd (1969) and McSween et al. (1988) reviewed metamorphism of the ordinary chondrites and emphasized several key points – metamorphism took place at temperatures of ~ 400 – 950 $^{o}$C, pressures less than ~ 2,000 bars, under dry conditions, was isochemical with respect to the major elements, and involved volatile element depletions with higher metamorphic grade. We used several of these conclusions as constraints on our calculations (e.g., temperature and pressure). However, we can also apply our results to questions such as the composition of metamorphic volatiles and the transport of volatile elements such as lead during metamorphism. We discuss the first topic below and are currently studying the latter topic, which is important for radiogenic dating of meteorites, to update our earlier work (Fegley 1990).

As our figures show, we predict that methane was the major carbon-bearing gas during thermal metamorphism of ordinary chondrites. However, in much of the asteroidal literature, CO is assumed to be the major carbon gas (e.g. Hashizume and Sugiura 1998, Krot et al. 1997, Wasson et al. 1993, Lee et al. 1992, Dodd 1969, Mueller 1964). The choice of CO gas may seem to be supported by heating experiments on chondrites, which typically produce much more CO than methane (e.g. Muenow et al. 1995). However, most of these experiments are done at very low pressures where CO would be the major C-bearing gas at equilibrium (see Figure 7a). For example, Muenow et al. (1995) heated several H and L chondrites in vacuum at ~$10^{-10}$ bars. As we verified for ourselves, this pressure is inside the CO stability field, but it is off scale to the right side of Figure 7a. However, for the pressures that should exist within a meteorite parent body, methane clearly predominates.

The myth that CO is the major C-bearing gas during meteorite metamorphism apparently dates back to Mueller (1964). However, his calculations only show the ratios of different gas pairs (e.g., $CO/CO_2$, $H_2/H_2O$) and are not complete chemical equilibrium calculations that simultaneously consider the dual constraints of mass balance and chemical equilibrium. Furthermore, he never computed $CH_4$/CO ratios! As far as we can tell, the study by Bukvic (1979) was the first time anyone computed gas chemistry considering mass balance and chemical equilibrium at P – T conditions like those for ordinary chondrite metamorphism.

*4.2 Planetary Out-gassing*

A major impetus for this work was a desire to understand the nature of outgassed planetary atmospheres, particularly that of the Earth. The terrestrial atmosphere is believed to be a product of outgassing of volatile-bearing material during and/or after the Earth's accretion (see e.g., Lange and Ahrens 1982, Lewis and Prinn 1984, Abe and Matsui 1985, Prinn and Fegley 1987, Kasting et al. 1993). However, the composition of the Earth's early atmosphere is controversial.

In the 1950s experiments by Stanley Miller and Harold Urey generated amino acids, carboxylic acids, hydrogen cyanide, and many other organic compounds by sparking a mixture of $CH_4$, $H_2$, $H_2O$, and $NH_3$ (Miller 1955, Miller and Urey 1959). Miller and Urey chose this reducing atmosphere because observations of Jupiter and Saturn showed that they contained ammonia and methane. Large amounts of $H_2$ were inferred to be present on Jupiter and Saturn The gas giant planet atmospheres were regarded as captured remnants of the solar nebula and the atmospheres of the early terrestrial planets were assumed by analogy to have been similar (e.g., Oparin 1938). Other scientists subsequently did many related experiments using electrical discharges, heat, or UV light applied to reducing gas mixtures (e.g., see Oro et al. 1990 and references therein). Miller-Urey reactions work best in $H_2$ + $CH_4$-bearing atmospheres, are still viable in mildly reducing atmospheres of $H_2$ + CO, but significantly less efficient in atmospheres of $H_2O$ + $CO_2$ (Stribling and Miller 1987).

However, atmospheric photochemists discovered that a reducing atmosphere of $CH_4$ and $NH_3$ was extremely vulnerable to destruction by UV sunlight (Kuhn and Atreya 1979; Kasting et al. 1983). Since the Sun's UV flux was higher in the past, it was believed that such an atmosphere could not exist for a sufficient length of time to support the origin of life. This was especially true because, before the rise of methanogenic bacteria, there was no known mechanism capable of generating large amounts of $CH_4$.

Research done in the 1980s and 1990s suggested that the Earth's early atmospheric composition must have been more oxidizing. Impact degassing of comets (Ahrens et al. 1989, Abe and Matsui 1985, Lange and Ahrens 1982) releases primarily $H_2O$ + $CO_2$, although recent work suggests that significant amounts of $CH_4$ may also be produced during impact degassing (Kress and McKay 2004). Work on the oxidation state of the mantle suggests that it has not changed significantly over time (since ~3900 Ma), which means that volcanic gases during the past were oxidizing, similar to those of the present ($H_2O$ + $CO_2$) (Delano 2001). All of these discoveries and lack of a source of reducing gases suggested that the earliest atmosphere must have been oxidizing ($H_2O$ + $CO_2$), or at least neutral ($H_2$ + CO), but definitely not reducing ($H_2$ + $CH_4$)

However the work by Bukvic (1979) and our results imply that Earth's early atmosphere was reducing. As mentioned earlier, Bukvic (1979) used a mixture of ~ 90% H- and 10% CI-chondrites to model volatile outgassing on the early Earth. He calculated the gas equilibrium composition within each layer of the planet's interior. Bukvic used a volcanic out-gassing scenario, assuming that the gas moved from depth to the surface without interacting or equilibrating with the intervening layers, to predict the composition of the initial atmosphere. He found that the major out-gassing products were $N_2$ at low temperatures and $CH_4$ and $H_2$ at higher temperatures. Figure 18 shows our calculations for average H-chondritic material along the terrestrial thermal profile used by Bukvic (1979). Our calculated compositions are nearly identical to his. Similar results were also obtained by Saxena and Fei (1988), who modeled mantle-fluid compositions for a chondritic system. Their results for P = 10 kilobar, T = 1273 K for a carbonaceous chondrite are shown in Fig. 18 as small symbols and are very similar to our results. Saxena and Frei (1988) restricted their results to deep mantle conditions. However, we find that similar results hold at shallow depths as well.

Recent developments suggest that a reducing atmosphere is more stable than previously believed. Tian et al. (2005) found that hydrogen escape from the Earth's atmosphere was less efficient than previously believed. This allows an atmosphere with a large H/C ratio to be sustained much longer. Observations of Titan's atmosphere, which is composed primarily of $CH_4$ and $N_2$, show that photochemically produced hydrocarbon aerosols form a haze layer in the upper atmosphere that protects the lower atmosphere from photochemical destruction. Such a haze layer could also have been produced on the early Earth from outgassed methane and ammonia (Zahnle 1986; Sagan and Chyba 1997; Pavlov et al. 2000). The reduced rate of hydrogen loss coupled with a UV-shield of organic haze particles could have protected our predicted $CH_4 + H_2$ atmosphere from rapid destruction. However, detailed study of volatile outgassing on the early Earth and the fate of the outgassed atmosphere is beyond the scope of this paper and has to be pursued elsewhere.

## 5. Summary

We modeled out-gassing of ordinary chondritic material in meteorite parent body and planetary settings. We found that for all three types of ordinary chondritic material (H, L, LL), the primary outgassed volatiles are $CH_4$ and $H_2$ gases. This holds true over a wide range of temperatures and pressures, and for variable volatile abundances. We found that the gas equilibrium composition was relatively insensitive to variations in temperature, pressure, volatile element abundances, and dissolution of C and N in iron metal. To first-order, an open system favored highly reduced gases $H_2$, $CH_4$, and $NH_3$ throughout the planetesimal. For a terrestrial planetary setting, we found that gases remained highly reducing, with $CH_4$ being the favored product. Our predicted atmospheric composition favors the Miller-Urey synthesis of organic compounds, which may have significant implications for the origin of life on Earth.

**Appendix 1.** List of solid phases included in our calculations.

| | | | | |
|---|---|---|---|---|
| **Element** | NaCl | $KNO_2$ | **Phosphate/phosphide** | $K_2Mg_2(SO_4)_3$ |
| Al | $NiCl_2$ | $KNO_3$ | $Ca_5(PO_4)_3Cl$ | $KAl(SO_4)_2$ |
| C(diamond) | $TiCl_2$ | $Mg(NO_3)_2$ | $Ca_5(PO_4)_3F$ | $KAl_3(OH)_6(SO_4)_2$ |
| C(graphite) | $TiCl_3$ | $Mg_3N_2$ | $Ca_5(PO_4)_3(OH)$ | $MgSO_4$ |
| Ca | $TiCl_4$(l) | $NaNO_2$ | $Fe_3P$ | $MnSO_4$ |
| Co | **Fluoride** | $NaNO_3$ | **Silicate** | $Na_2SO_4$ |
| Cr | $AlF_3$ | $Si_3N_4$ | $Al_2Si_4O_{10}(OH)_2$ | $NiSO_4$ |
| Fe | $CaF_2$ | $Si_2N_2O$ | $Al_6Si_2O_{13}$ | **sulfide** |
| K | $CoF_2$ | TiN | $Al_2Si_2O_5(OH)_4$ | $Al_2S_3$ |
| Mg | $CoF_3$ | **oxide** | $Ca_2Al_2Si_3O_{10}(OH)_2$ | CaS |
| Mn | $CrF_2$ | $Al_2O_3$ | $Ca_2Al_2SiO_7$ | $Co_3S_4$(L) |
| Na | $CrF_3$ | CaO | $Ca_2MgSi_2O_7$ | $Co_9S_8$(L) |
| Ni | $FeF_2$ | $CaAl_{12}O_{19}$ | $CaAl_2Si_2O_8$ | CoS |
| P | $FeF_3$ | $CaAl_4O_7$ | $CaAl_4Si_2O_{10}(OH)_2$ | $CoS_2$ |
| S | KF | $CaTiO_3$ | $CaFeSi_2O6$ | CrS |
| Si | $K_3AlF_6$ | CoO | $CaSiO_3$ | $Fe_{0.875}S$ |
| Ti | $MgF_2$ | $Co_3O_4$ | $Fe_7Si_8O_{22}(OH)_2$ | FeS |
| **Carbide** | $MnF_2$ | $Cr_2O_3$ | $Fe_2SiO_4$ | $FeS_2$ |
| $Al_4C_3$ | $MnF_3$ | $Fe_{0.947}O$ | $FeSiO_3$ | $K_2S$ |
| $Cr_{23}C_6$ | $MnF_4$ | $Fe_3O_4$ | $KAl_3Si_3O_{10}(OH)_2$ | MgS |
| $Cr_3C_2$ | NaF | $Fe_2O_3$ | $KAlSi_3O_8$ | MnS |
| $Cr_7C_3$ | $Na_3AlF_6$ | FeO | $K_2Si_2O_4$ | $MnS_2$ |
| $Fe_3C$ | $Na_5Al_3F_{14}$ | $FeCr_2O_4$ | $K_2SiO_3$ | $Na_2S$ |
| $Fe_2C$ | $NiF_2$ | $FeTiO_3$ | $Mg_2SiO_4$ | $Ni_3S_2$ |
| $Mn_{15}C_4$ | $TiF_2$ | $H_2O$ | $Mg_3Si_2O_5(OH)_4$ | NiS |
| $Mn_{23}C_6$ | $TiF_3$ | $H_2O_2$ | $Mg_3Si_4O_{10}(OH)_2$ | $NiS_2$ |
| $Mn_3C$ | $TiF_4$ | $K_2O$ | $Mg_7Si_8O_{22}(OH)_2$ | SiS |
| $Mn_5C_3$ | **hydride** | $K_2O_2$ | $MgCaSi_2O_6$ | $SiS_2$ |
| $Mn_7C_3$ | $AlH_3$ | $KO_2$ | $MgSiO_3$ | TiS |
| SiC | $CaH_2$ | $KAlO_2$ | $Mn_2SiO_4$ | $TiS_2$ |
| TiC(L) | KH | MgO | $MnSiO_3$ | |
| **carbonate** | $MgH_2$ | $MgAl_2O_4$ | $Na_8Al_6Si_6O_{24}Cl_2$ | |
| $CaCO_3$ | NaH | $Mn_2O_3$ | $NaAl_3Si_3O_{10}(OH)_2$ | |
| $CaMg(CO_3)_2$ | **Hydroxide** | $Mn_2O_7$ (L) | $NaFeSi_2O_6$ | |
| $FeCO_3$ | $Al(OH)_3$ | $Mn_3O_4$ | $NaAlSi_3O_8$ | |
| $K_2CO_3$ | $Ca(OH)_2$ | MnO | $NaAlSiO_4$ | |
| $MgCO_3$ | $Co(OH)_2$ | $MnO_2$ | $Na_2Si_2O5$ | |
| $MnCO_3$ | $Fe(OH)_2$ | $Na_2O$ | $Na_2SiO_3$ | |
| $Na_2CO_3$ | $Fe(OH)_3$ | $Na_2O_2$ | **silicide** | |
| $NaAlCO_3(OH)_2$ | FeOOH | $NaAlO_2$ | $Cr_3Si$ | |
| **chloride** | KOH | $NaO_2$ | $Cr_5Si_3$ | |
| $AlCl_3$ | $Mg(OH)_2$ | NiO | CrSi | |
| $CaCl_2$ | $Mn(OH)_2$ | $SiO_2$(trid) | $CrSi_2$ | |
| $CoCl_2$ | MnOOH | $SiO_2$(gl) | $Ti_5Si_3$ | |
| $CrCl_2$ | NaOH | $SiO_2$(crist) | TiSi | |
| $CrCl_3$ | $Ni(OH)_2$ | $SiO_2$(QTz) | $TiSi_2$ | |
| $FeCl_2$ | NiOOH | $Ti_2O_3$ | **Sulfate** | |
| $FeCl_3$ | **Nitride/nitrate** | $Ti_3O_5$ | $Al_2(SO_4)_3$ | |
| FeOCl | AlN | $Ti_4O_7$ | $CaSO_4$ | |
| KCl | CrN | TiO | $CaSO_4 \cdot 2H_2O$ | |
| $MgCl_2$ | $Cr_2N$ | $TiO_2$ | $Fe_2(SO_4)_3$ | |
| $MnCl_2$ | $Fe_4N$ | | $K_2SO_4$ | |

**Acknowledgments**.

This work was supported by the NASA Astrobiology and Origins Programs. We thank K. Lodders for her advice.

Table 1. Composition of average ordinary chondrite falls

| | Mass % of component | | | | | |
|---|---|---|---|---|---|---|
| | H-chondrite[a] | | L-chondrite[b] | | LL-chondrite[c] | |
| Component | Range | Average | Range | Average | Range | Average |
| $SiO_2$ | 30.07 – 39.73 | 36.49 | 35.43 – 42.00 | 39.66 | 37.1 – 41.80 | 40.48 |
| $TiO_2$ | 0.05 – 0.45 | 0.12 | 0.02 – 0.18 | 0.12 | 0.08 – 0.23 | 0.13 |
| $Al_2O_3$ | 1.46 – 4.48 | 2.35 | 1.35 – 4.26 | 2.32 | 1.84 – 3.91 | 2.38 |
| $Cr_2O_3$ | 0.10 – 0.73 | 0.47 | 0.05 – 0.76 | 0.50 | 0.36 – 0.72 | 0.55 |
| FeO | 4.74 – 17.54 | 9.54 | 10.74 – 17.61 | 14.14 | 11.46 – 21.11 | 17.44 |
| MnO | 0.21 – 0.50 | 0.30 | 0.07 – 1.21 | 0.34 | 0.05 – 0.45 | 0.33 |
| MgO | 19.77 – 25.01 | 23.41 | 21.11 – 26.6 | 24.84 | 23.81 – 26.50 | 25.32 |
| CaO | 1.18 – 2.66 | 1.70 | 1.41 – 2.48 | 1.84 | 1.47 – 2.24 | 1.82 |
| $Na_2O$ | 0.35 – 1.29 | 0.87 | 0.50 – 2.29 | 0.98 | 0.75 – 1.21 | 0.96 |
| $K_2O$ | 0.04 – 0.23 | 0.10 | 0.03 – 0.38 | 0.12 | 0.05 – 0.16 | 0.11 |
| $P_2O_5$ | 0 – 0.53 | 0.26 | 0.06 – 0.68 | 0.25 | 0.13 – 0.65 | 0.27 |
| Fe (m) | 8.1 – 26.19 | 16.72 | 2.78 – 14.68 | 7.27 | 0.33 – 12.93 | 2.64 |
| Ni | 1.15 – 2.24 | 1.70 | 0.70 – 1.67 | 1.23 | 0.73 – 1.39 | 1.03 |
| Co | 0.01 – 0.17 | 0.08 | 0.01 – 0.14 | 0.06 | 0.02 – 0.09 | 0.054 |
| FeS | 2.52 – 9.65 | 5.33 | 3.45 – 9.88 | 5.99 | 4.03 – 8.39 | 5.87 |
| $H_2O^+$ | 0 – 0.92 | 0.32 | 0 – 1.42 | 0.34 | 0 – 1.77 | 0.60 |
| $H_2O^-$ | 0 – 0.36 | 0.09 | 0 – 0.36 | 0.07 | 0 – 0.47 | 0.16 |
| C | 0.01 – 0.34 | 0.12 | 0.01 – 0.90 | 0.16 | 0.02 – 0.57 | 0.235 |
| N(ppm) | 2.05 – 121 | 34 | 4.02 – 109 | 34 | 3 – 298 | 50 |
| Cl (ppm) | 7 – 210 | 77 | 15 – 212 | 76 | 121 – 131 | 126 |
| F(ppm) | 8 – 41 | 27 | 8 – 52 | 28 | 49 – 66 | 58 |
| Total | – | 99.98 | – | 100.24 | – | 100.40 |

[a]No. of analyses: oxides and metals (90), water (47), C (47), N (60), Cl (30), F(7)
[b]No. of analyses: oxides and metals (123), water (84), C (66), N (88), Cl (22), F(11)
[c]No. of analyses: oxides and metals (29), water (18), C (19), N (39), Cl (2), F(3)

Table 2. Solid Solution compositions

| | H chondrites | |
|---|---|---|
| Solid solution | observed | Hebe (3.7 Ma)[a] |
| Olivine | $Fo_{80-84}Fa_{20-16}$ | $Fo_{77-86}Fa_{23-14}$ |
| Ca-poor pyroxene[b] | $En_{86-80}Fs_{14-20}$ | $En_{83-67}Fs_{17-28}Wo_{0-5}$ |
| Plagioclase | $An_{12}Ab_{82}Or_{6}$ | $An_{21-39}Ab_{73-50}Or_{6-11}$ |
| | **L chondrites** | |
| Solid solution | observed | Hebe (3.7 Ma)[a] |
| Olivine | $Fo_{75-79}Fa_{25-21}$ | $Fo_{71-80}Fa_{29-20}$ |
| Ca-poor pyroxene[b] | $En_{81-78}Fs_{19-22}$ | $En_{79-60}Fs_{19-31}Wo_{0-9}$ |
| Plagioclase | $An_{10}Ab_{84}Or_{6}$ | $An_{14-40}Ab_{79-42}Or_{6-18}$ |
| | **LL chondrites** | |
| Solid solution | Observed | Hebe (3.7 Ma)[a] |
| Olivine | $Fo_{68-74}Fa_{32-26}$ | $Fo_{67-75}Fa_{33-25}$ |
| Ca-poor pyroxene[b] | $En_{78-74}Fs_{22-26}$ | $En_{77-56}Fs_{21-34}Wo_{0-9}$ |
| Plagioclase | $An_{10}Ab_{86}Or_{4}$ | $An_{17-47}Ab_{77-38}Or_{6-16}$ |

[a]Calculated solid solution composition for the 3.7 Ma Hebe T/P profile from low T to high T (e.g. olv = $Fo_{77}Fa_{23}$ @ 300 K, olv = $Fo_{86}Fa_{14}$ @ 1225 K for H-chondrites).

[b]For calculated compositions, remainder is found in $MnSiO_3$

Table 3. Normative mineralogy of ordinary chondrites compared with calculated mineralogies

| Mineral (mass %) | Avg. H chondrite[a] | Hebe (3.7 Ma)[b] |
|---|---|---|
| Apatite | 0.65±0.07 | 0.62 |
| Chromite | 0.76±0.05 | 0.70 |
| Ilmenite | 0.23±0.02 | 0.23 |
| Orthoclase | 0.56±0.07 | 0.60 |
| Albite | 7.30±0.31 | 7.33-2.72 |
| Anorthite | 1.70±0.43 | 2.22 – 2.27 |
| Diopside | 4.11±0.54 | 3.48 – 0 |
| Hypersthene | 26.15±2.94 | 22.18 – 35.63 |
| Olivine | 35.04±3.93 | 39.10 – 31.90 |
| Metal | 18.02±1.65 | 17.36 – 16.78 |
| FeS | 5.47±0.38 | 4.96 |
| total | 99.99 | 98.78 – 96.41[c] |

| Mineral (wt%) | Avg. L chondrite[a] | Hebe (3.7 Ma)[b] |
|---|---|---|
| Apatite | 0.54±0.09 | 0.59 |
| Chromite | 0.78±0.05 | 0.74 |
| Ilmenite | 0.24±0.02 | 0.23 |
| Orthoclase | 0.64±0.07 | 0.71 |
| Albite | 8.07±0.44 | 8.29 – 1.51 |
| Anorthite | 1.59±0.49 | 1.58 – 1.55 |
| Diopside | 4.97±0.66 | 4.59 – 0 |
| Hypersthene | 24.15±2.59 | 20.66 – 38.44 |
| Olivine | 44.83±3.02 | 49.32 – 38.39 |
| Metal | 8.39±0.99 | 7.30 – 6.59 |
| FeS | 5.80±0.80 | 5.99 |
| total | 100.00 | 100.00 – 92.64[c] |

| Mineral (wt%) | Avg. LL chondrite[a] | Hebe (3.7 Ma)[b] |
|---|---|---|
| Apatite | 0.54±0.11 | 0.64 |
| Chromite | 0.80±0.04 | 0.81 |
| Ilmenite | 0.25±0.04 | 0.25 |
| Orthoclase | 0.61±0.12 | 0.65 |
| Albite | 8.11±0.50 | 8.13 – 1.45 |
| Anorthite | 1.55±0.40 | 1.86 – 1.91 |
| Diopside | 5.34±0.54 | 4.19 – 0 |
| Hypersthene | 21.44±3.73 | 17.10 – 35.83 |
| Olivine | 51.92±5.57 | 58.60 – 46.04 |
| Metal | 3.59±1.65 | 1.60 – 2.41 |
| FeS | 5.85±1.06 | 5.16 – 5.89 |
| total | 100.00 | 98.99 – 95.88[c] |

[a]From McSween et al. (1991)
[b] calculated for the 3.7 Ma Hebe T/P profile. Ranges are from low to high T (300 – 1225 K).
[c] Other phases present: H: corundum, talc, $Fe_2C$, MnS, sodalite, nepheline; L: C, $Fe_2C$, sodalite, nepheline; LL: C, $Fe_2C$, sodalite, nepheline, $Ni_3S_2$

Figure Captions

Figure 1. Pressure vs. temperature profiles for asteroid 6 Hebe based on the temperature vs. depth profiles from Ghosh et al. (2003). The different lines refer to temperature profiles at 3.7, 5.7 and 7.3 Ma after accretion.

Figure 2. The chemical equilibrium gas composition as a function of temperature for the Holbrook (L6) meteorite at a total pressure of $10^{-3}$ bar.

Figure 3. Chemical equilibrium composition of volatiles outgassed from average H-chondritic material as a function of temperature for three different T – P profiles for the asteroid 6 Hebe. (a) 3.7 Ma, (b) 5.7 Ma, (c) 7.3 Ma after accretion. The thermal profiles are given in Fig. 1.

Figure 4. Chemical equilibrium composition of volatiles outgassed from average L-chondritic material as a function of temperature. The 3.7 Ma thermal profile for the asteroid 6 Hebe is used for the calculations.

Figure 5. Chemical equilibrium composition of volatiles outgassed from average LL-chondritic material as a function of temperature. The 3.7 Ma thermal profile for the asteroid 6 Hebe is used for the calculations.

Figure 6 Chemical equilibrium distribution of volatile elements between the major solid and gas phases for average H-chondritic material. Calculations use the Hebe 3.7 Ma

thermal profile. Results are expressed as percentage of each element as a function of temperature. (a) hydrogen, (b) carbon, (c) oxygen, (d) sulfur.

Figure. 7 (a) The chemical equilibrium distribution of carbon between $CH_4$, CO, and graphite for average H-chondritic material plotted as a function of temperature and pressure. The solid lines are equal abundance lines showing where $CH_4/CO = 1$, $CH_4$/graphite = 1, and graphite/CO = 1. The dashed line is the extension of the $CH_4/CO$ equal abundance line into the field where graphite is dominant. (b) The chemical equilibrium distribution of nitrogen between $N_2$ and $NH_3$ for average H-chondritic material plotted as a function of temperature and pressure. The solid lines are equal abundance lines showing where $N_2 = NH_3$. The region where $NH_3 > N_2$ is smaller for L- and LL- chondritic material. The dotted line shows where talc (in average H-chondritic material) or chlorapatite (in average L- and LL-chondritic material) is stable. The shaded region in both plots is the T – P range in thermal models for asteroid 6 Hebe (see Fig. 1).

Figure 8. (a) The chemical equilibrium oxygen fugacity of average H-chondritic material as a function of temperature compared to different solid-state oxygen fugacity buffers. The points are intrinsic oxygen fugacities of the Guareña (H6) and Ochansk (H4) chondrites measured by Brett and Sato (1984). Also shown are the intrinsic oxygen fugacities of the Guareña (H6) chondrite measured by Walter and Doan (1969) and the calculated oxygen fugacities for average H4-H6 chondrites from McSween and Lobotka (1993). (b) The chemical equilibrium oxygen fugacity of average L-chondritic material as a function of temperature compared to different solid-state oxygen fugacity buffers. The points are intrinsic oxygen fugacities of the Farmington (L5) chondrite measured by Brett and Sato (1984). Also shown are the intrinsic oxygen fugacities of the Holbrook (L6) chondrite measured by Walter and Doan (1969) and the calculated oxygen fugacities for average L4-L6 chondrites from McSween and Lobotka (1993). (c) The chemical equilibrium oxygen fugacity of average LL-chondritic material as a function of temperature compared to different solid-state oxygen fugacity buffers. The points are intrinsic oxygen fugacities of the Semarkona (LL3) and Cherokee Springs (LL5) chondrites measured by Brett and Sato (1984). (IW = iron-wüstite, QFI = quartz-fayalite-iron, QFM = quartz-fayalite-magnetite, WM = wüstite-magnetite)

Figure 9. The mineralogy of average H-chondritic material plotted as a function of temperature and pressure along the Hebe 3.7 Ma thermal profile. The calculations assume ideal solid solutions.

Figure 10. The equilibrium gas chemistry of average H-chondritic material plotted as a function of pressure at constant temperatures of 600 K, 900 K, and 1200 K.

Figure 11. The chemical equilibrium oxygen fugacity of average H-chondritic material as a function of temperature for pressures of 1 – 10,000 bars.

Figure 12. The equilibrium gas chemistry plotted as a function of temperature for average H-chondritic material in an open system. Calculations are done using the 3.7 Ma Hebe thermal profile.

Figure 13. Equilibrium gas chemistry as a function of temperature and hydrogen elemental abundance for average H-chondritic material: (a) H = 2.8 mg/g (2.5% $H_2O$), (b) H = 0.46 mg/g (0.41% $H_2O$ nominal value), (c) H = 0.01 mg/g (0.0089% $H_2O$), (d) H = 1 μg/g (8.9 μg/g $H_2O$). Calculations are done using the 3.7 Ma Hebe thermal profile.

Figure 14. Equilibrium gas chemistry as a function of temperature and carbon elemental abundance for average H-chondritic material: (a) C= 3.4 mg/g, (b) C = 1.2 mg/g (nominal value), (c) C = 0.1 mg/g. Calculations are done using the 3.7 Ma Hebe thermal profile.

Figure 15. Equilibrium gas chemistry as a function of temperature and nitrogen elemental abundance for average H-chondritic material: (a) N = 121 ppm, (b) N = 34 ppm (nominal value), (c) N = 5 ppm. Calculations are done using the 3.7 Ma Hebe thermal profile.

Figure 16. Equilibrium gas chemistry as a function of temperature and oxygen elemental abundance for average H-chondritic material: (a) O = 38 wt%, (b) O = 36 wt%, (c) O = 32 wt%, (d) O = 30 wt%, (e) O = 28 wt%. Calculations are done using the 3.7 Ma Hebe thermal profile.

Figure 17. (a) The chemical equilibrium distribution of nitrogen between gas and metal in average H-chondritic material plotted versus temperature. (b) The chemical equilibrium distribution of carbon between gas and metal in average H-chondritic material plotted versus temperature. Both calculations use the Hebe 3.7 Ma thermal profile.

Figure 18. Equilibrium gas chemistry as a function of temperature for average H-chondritic material for a terrestrial thermal profile.

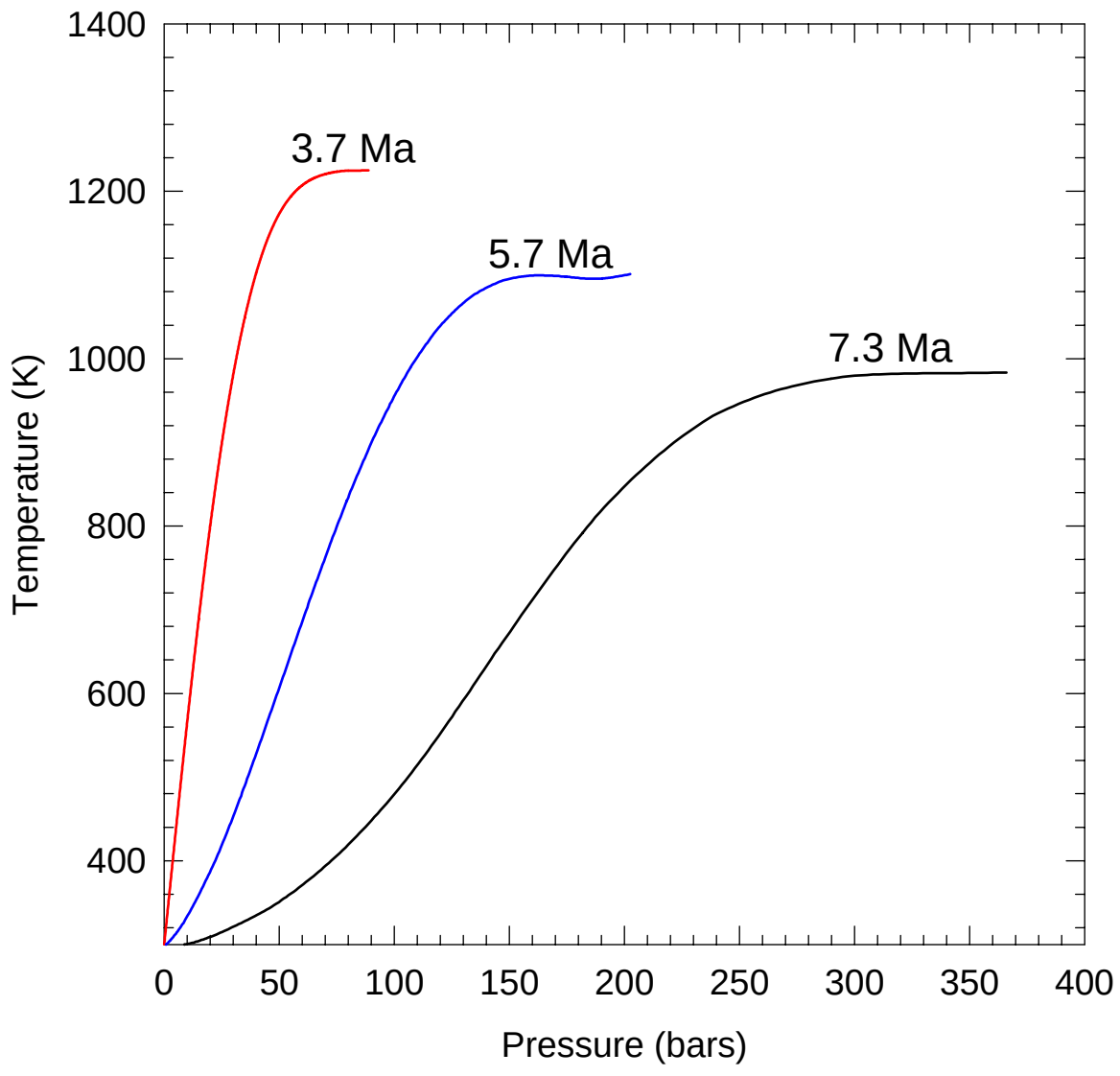

Figure 1. Schaefer and Fegley 2006

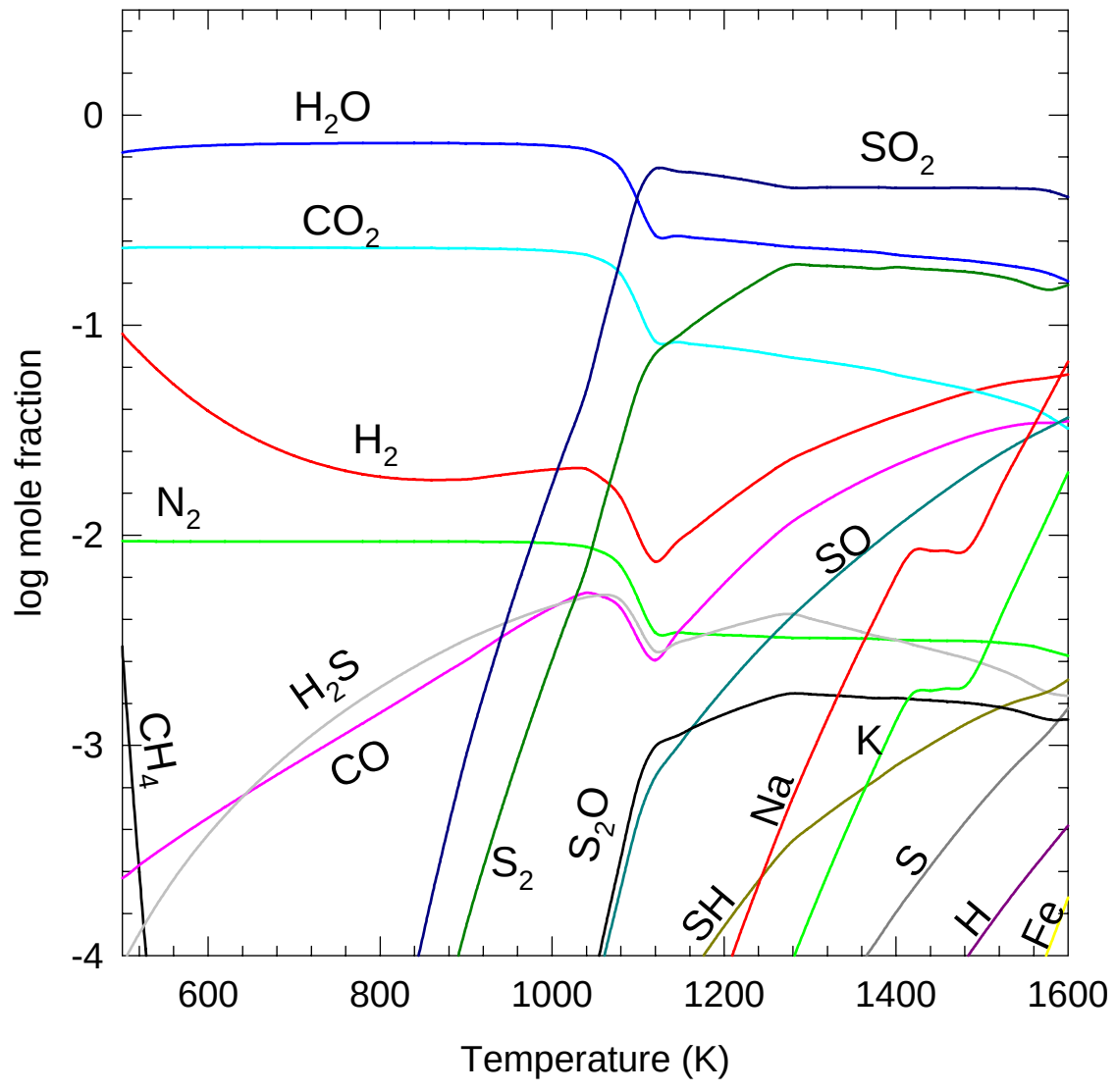

Figure 2. Schaefer and Fegley 2006

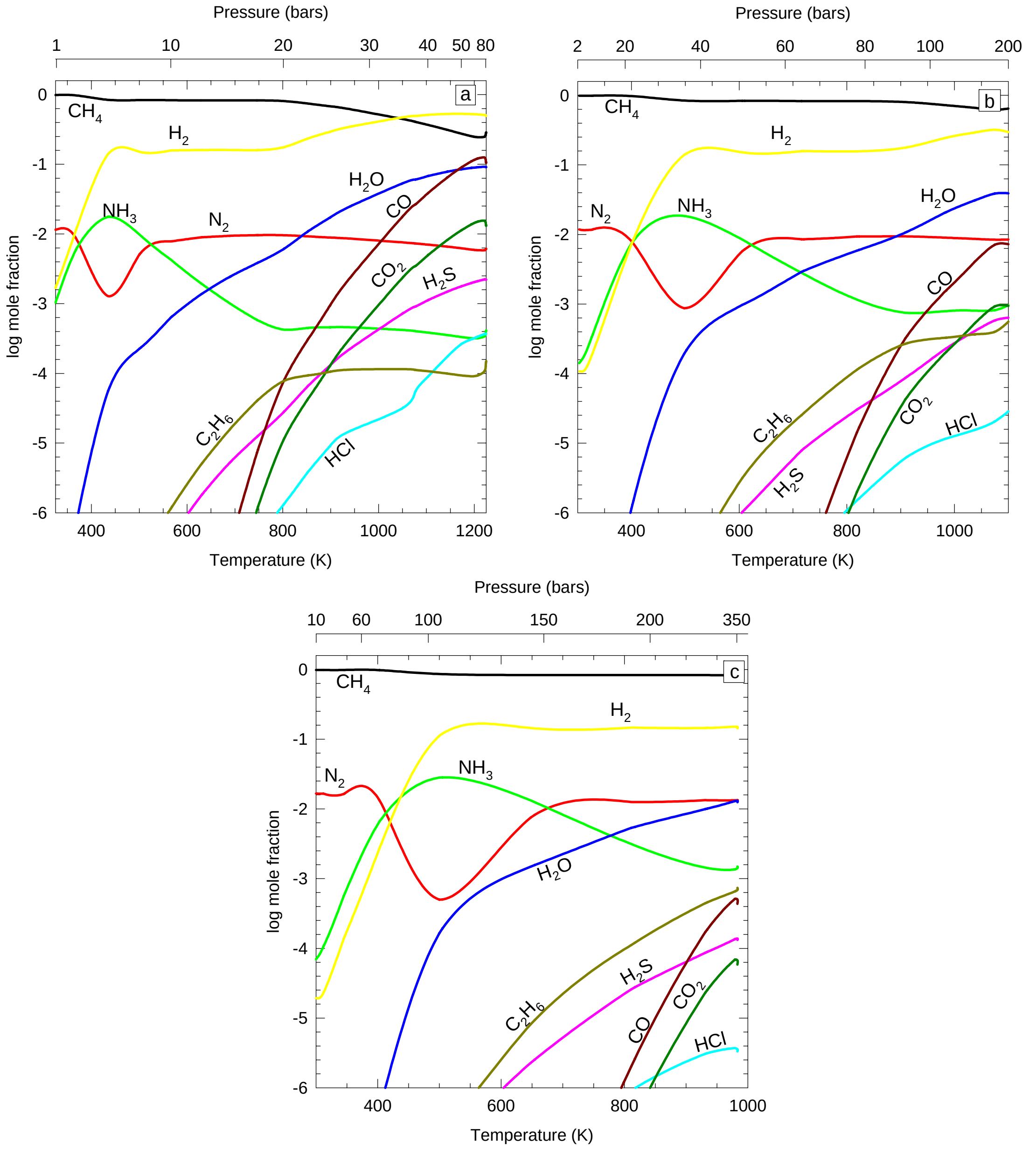

Figure 3. Schaefer and Fegley 2006.

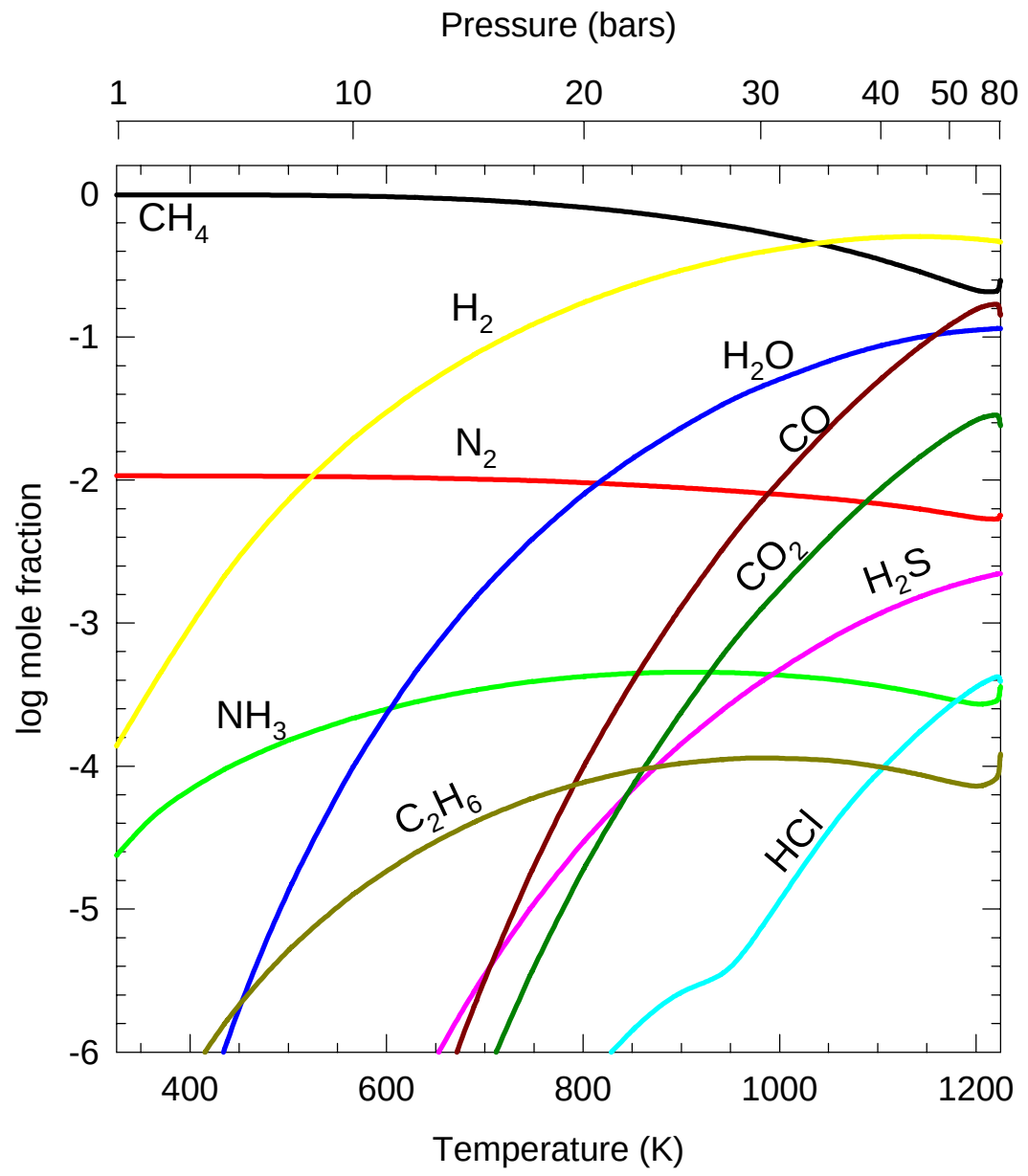

Figure 4. Schaefer and Fegley 2006

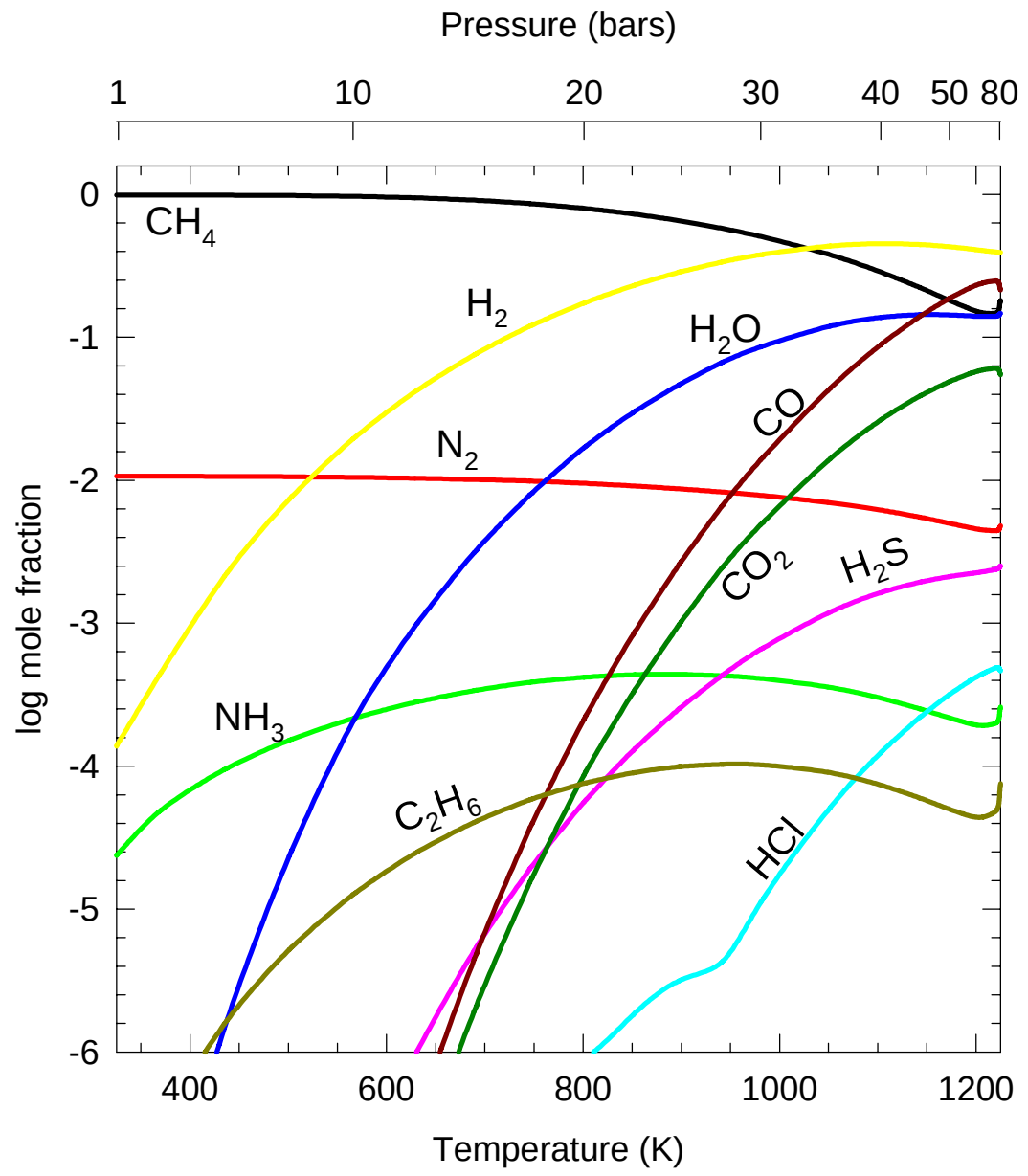

Figure 5. Schaefer and Fegley 2006

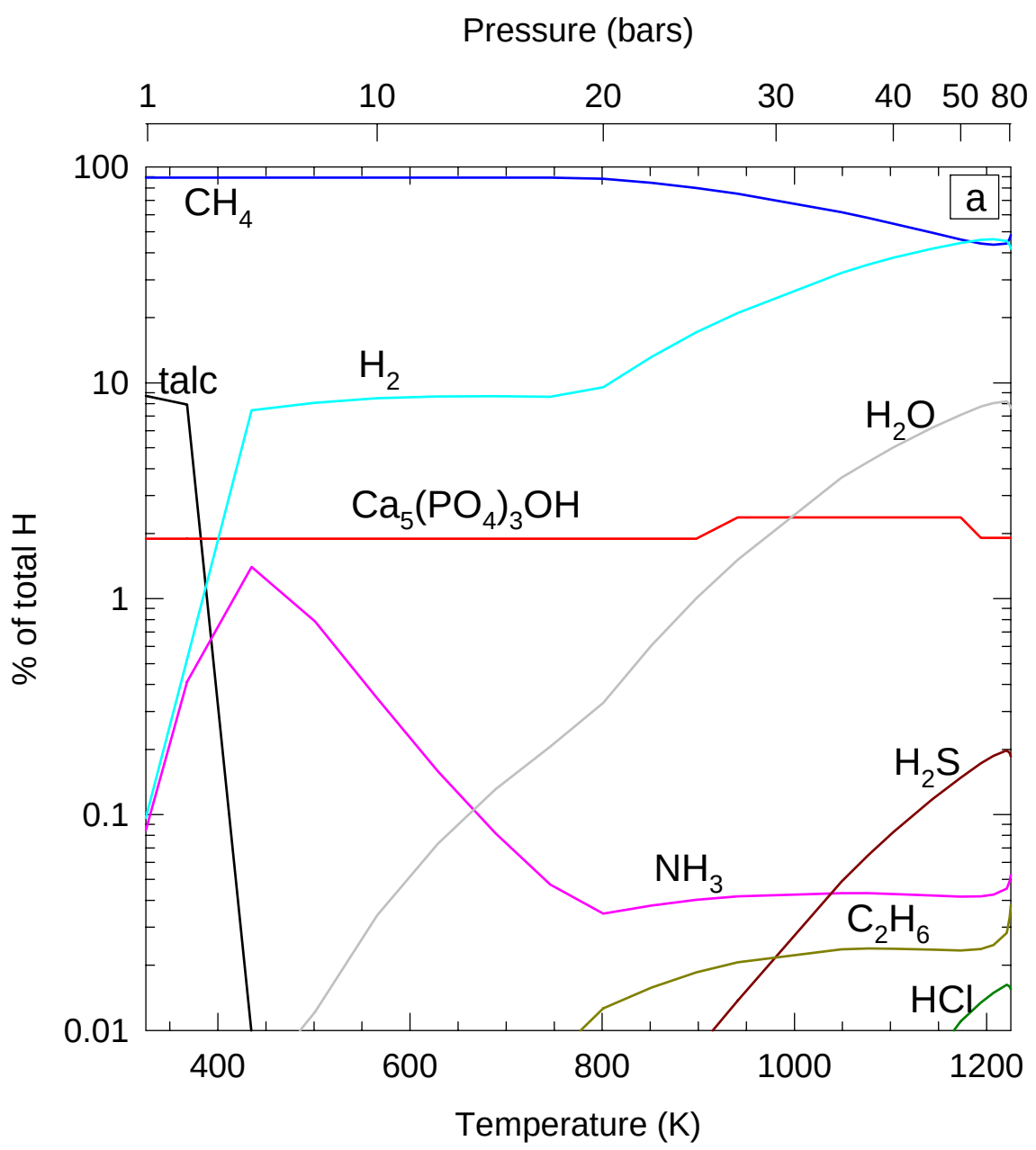

Figure 6(a) Schaefer and Fegley 2006

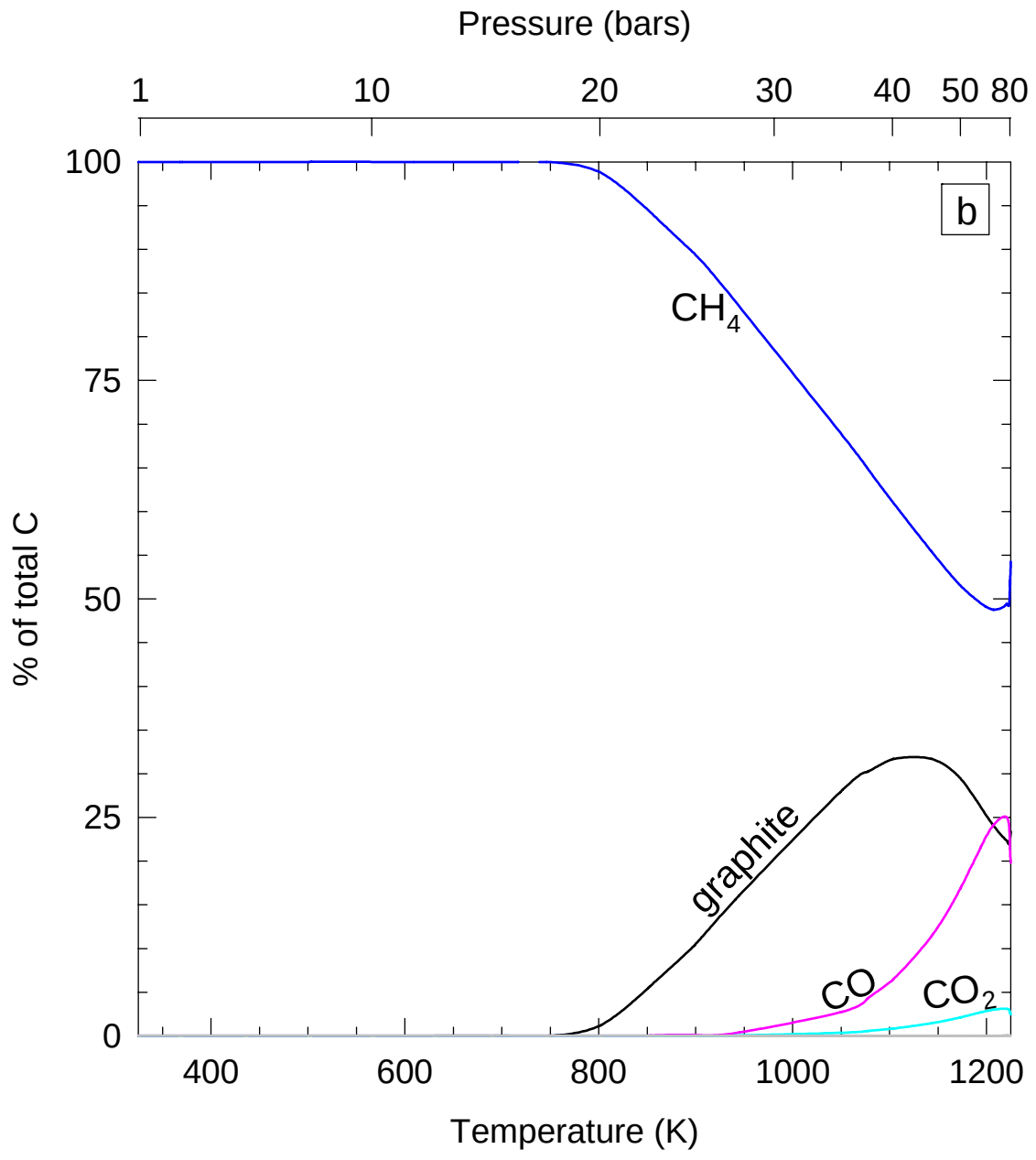

Figure 6(b) Schaefer and Fegley 2006

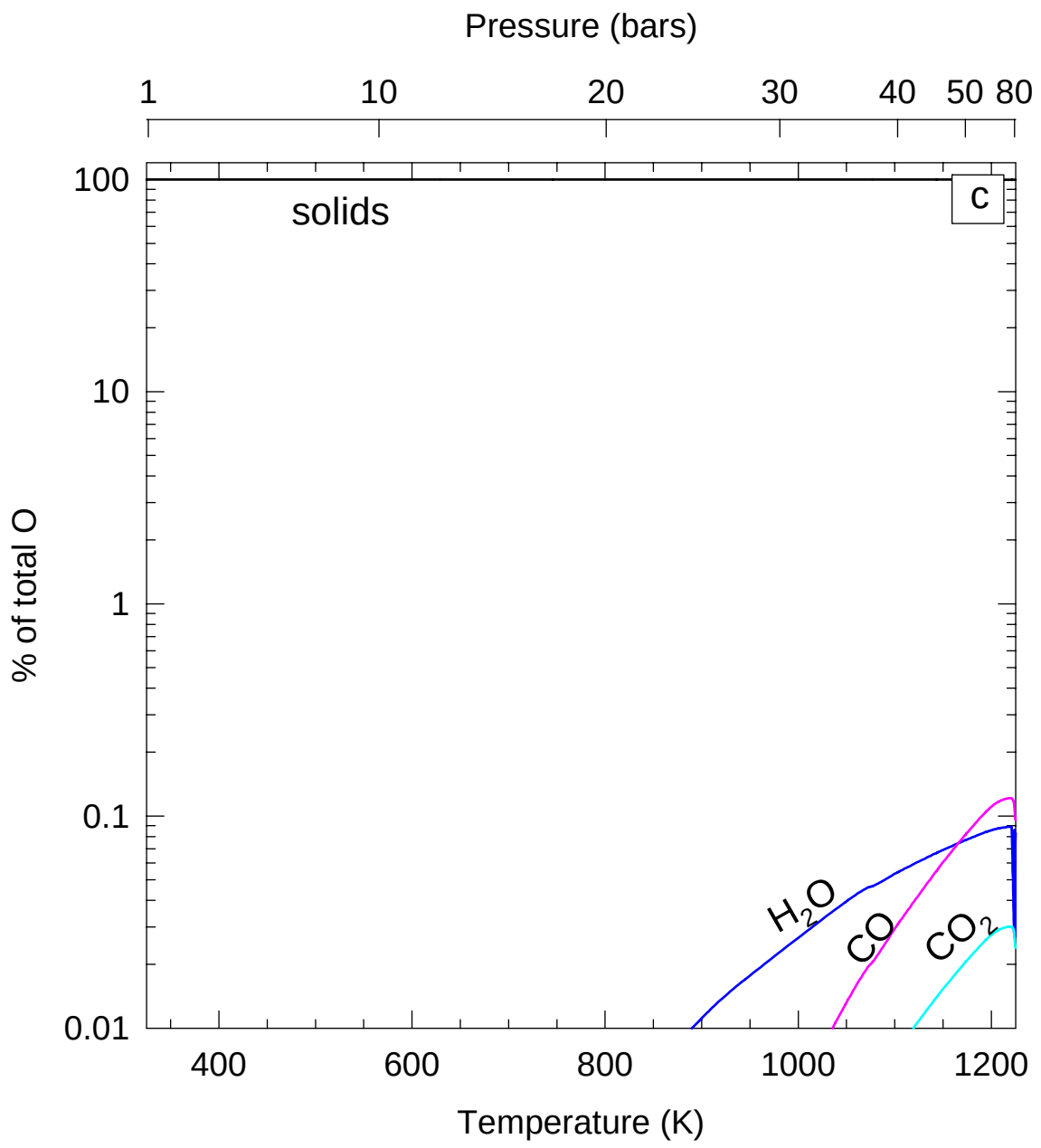

Figure 6(c) Schaefer and Fegley 2006

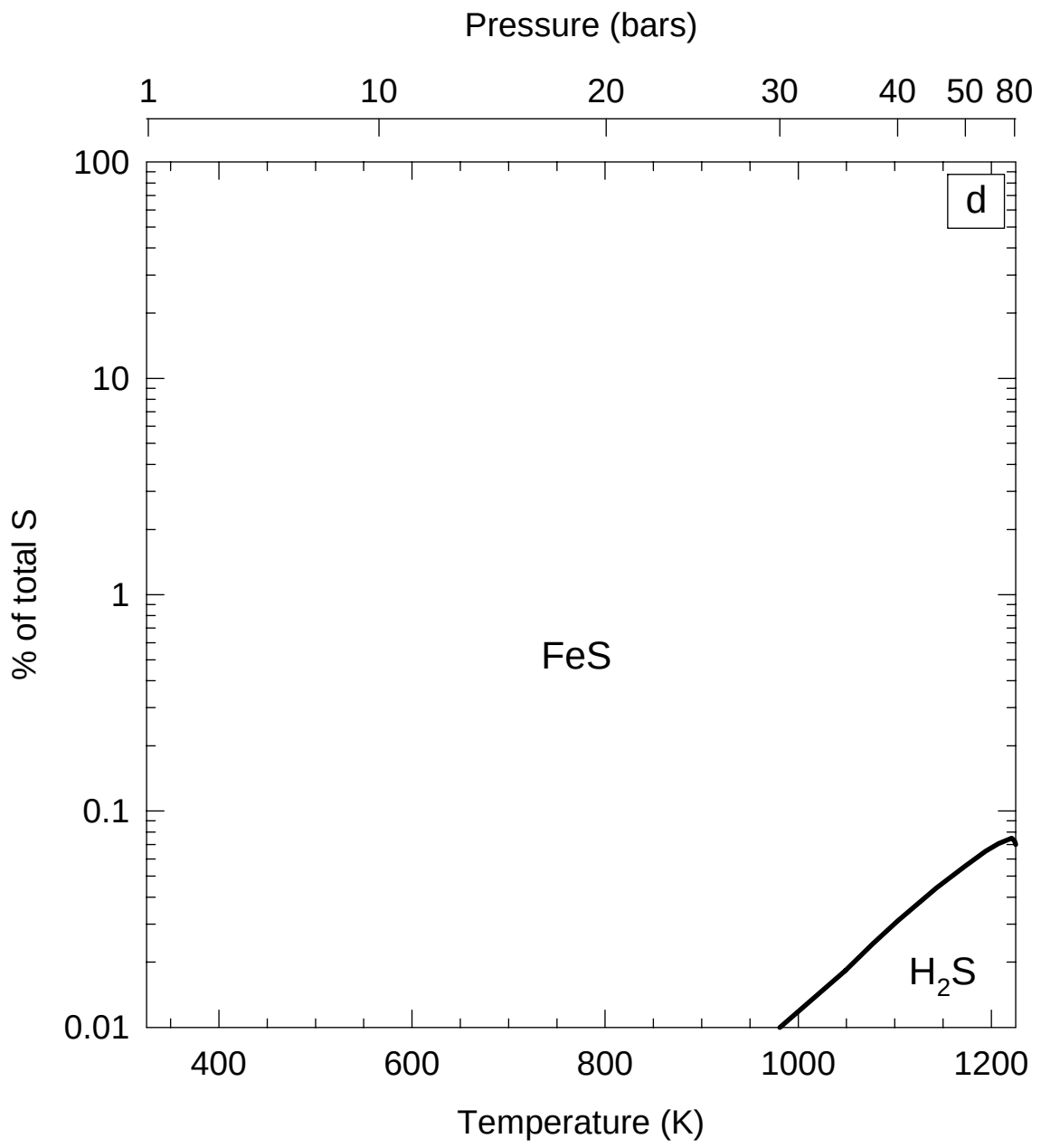

Figure 6(d) Schaefer and Fegley 2006

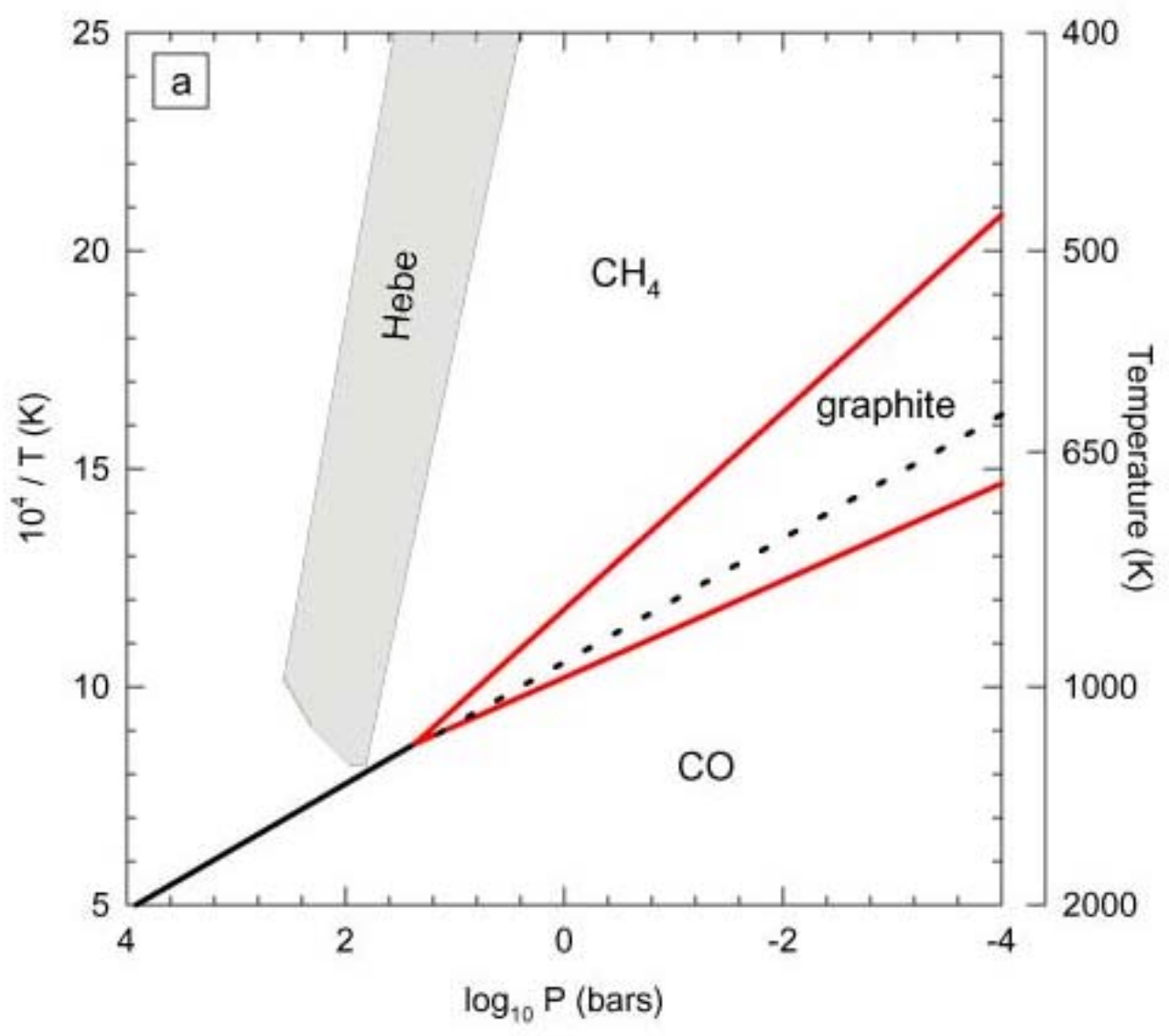

Figure 7 (a)

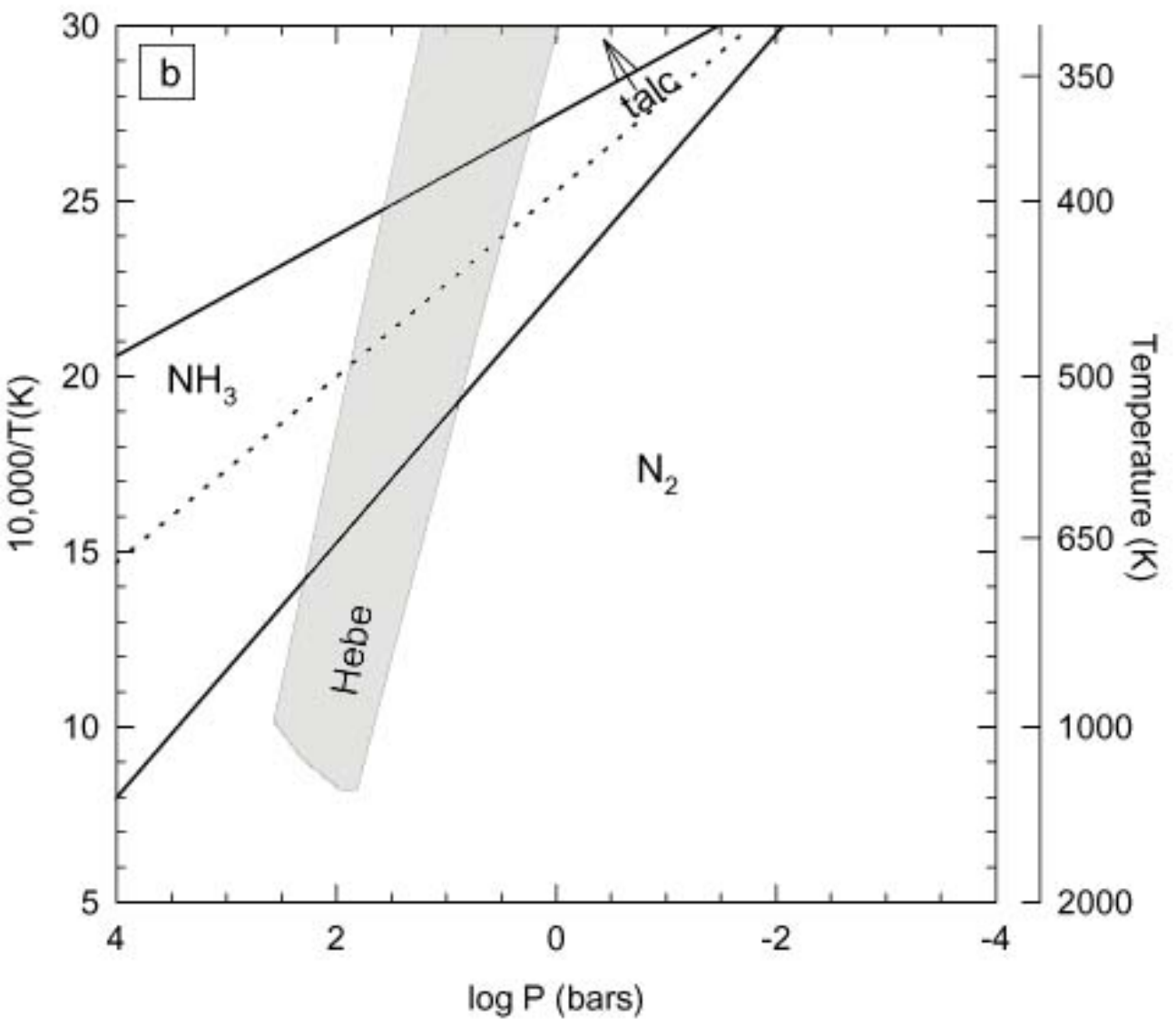

Figure 7(b) Schaefer and Fegley (2006)

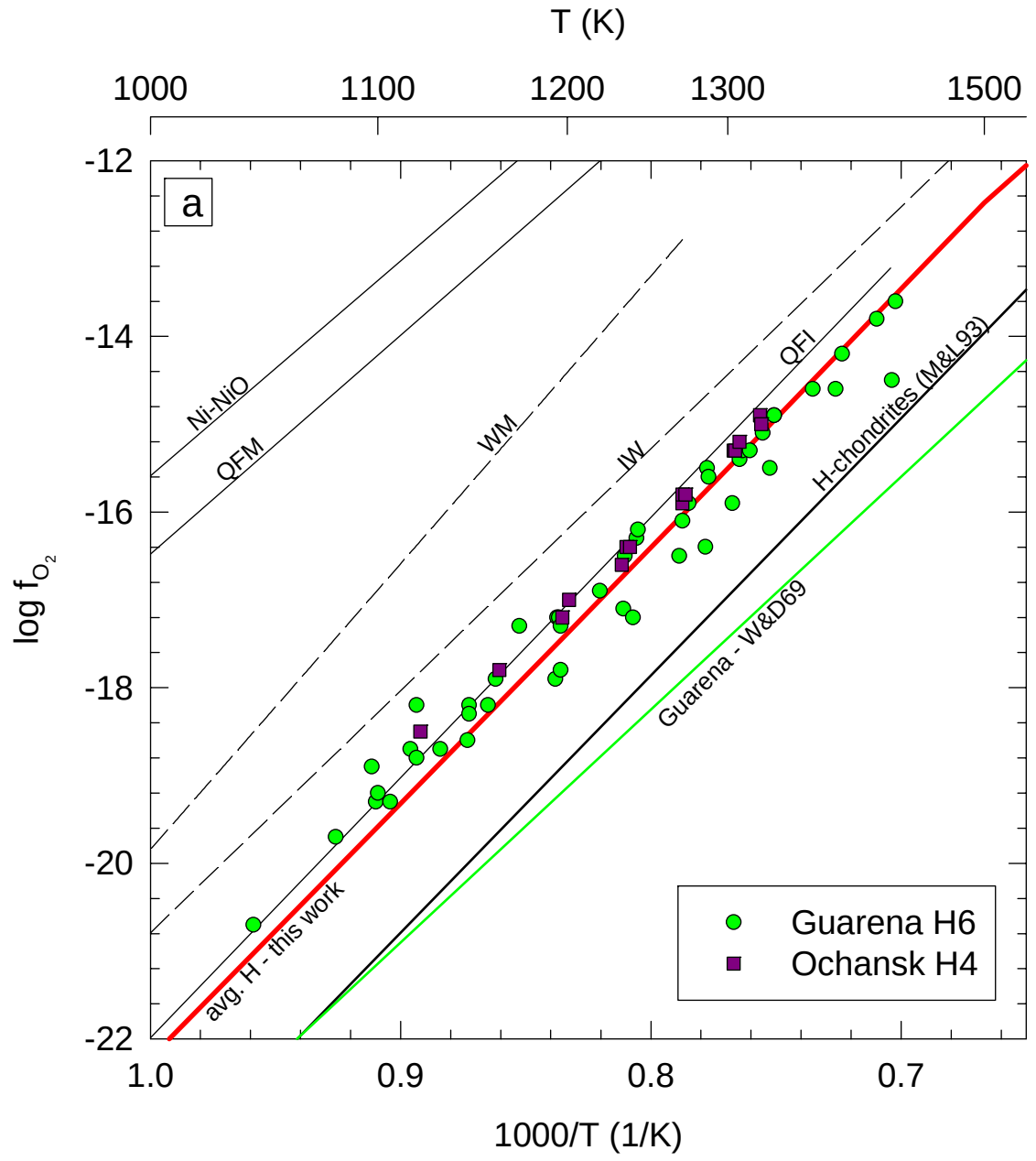

Figure 8(a). Schaefer and Fegley, 2006.

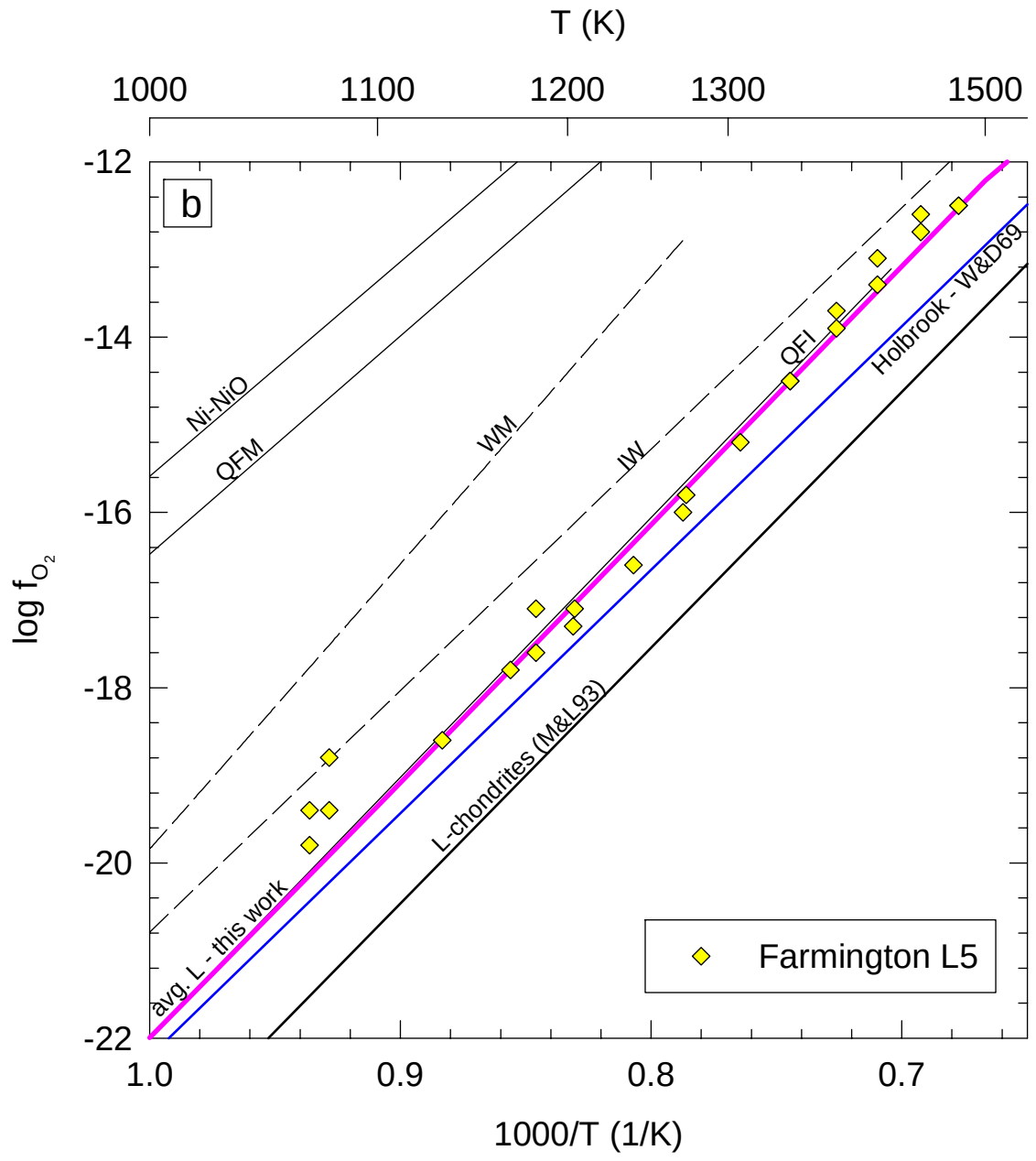

Figure 8(b) Schaefer and Fegley, 2006

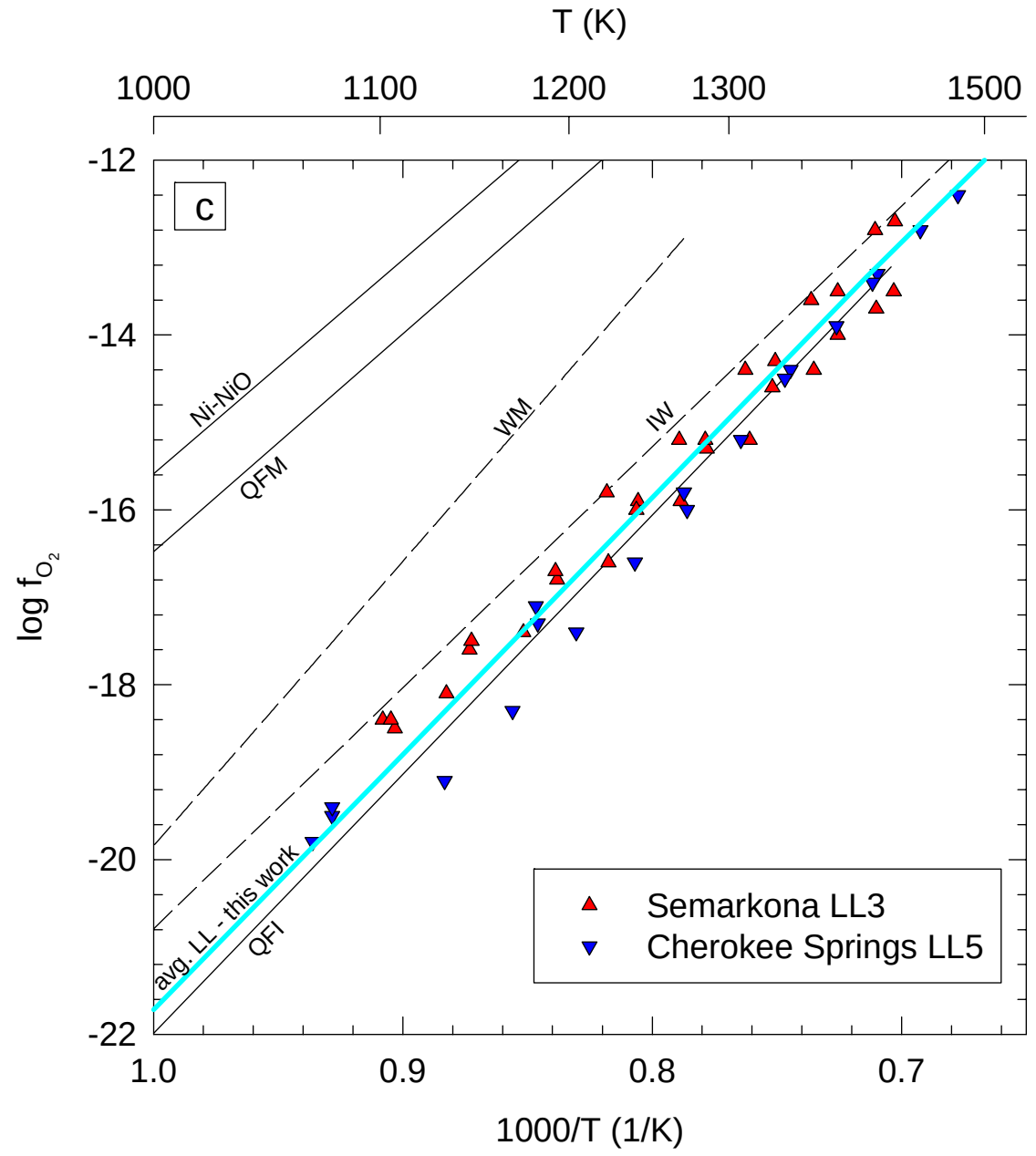

Figure 8(c), Schaefer and Fegley 2006

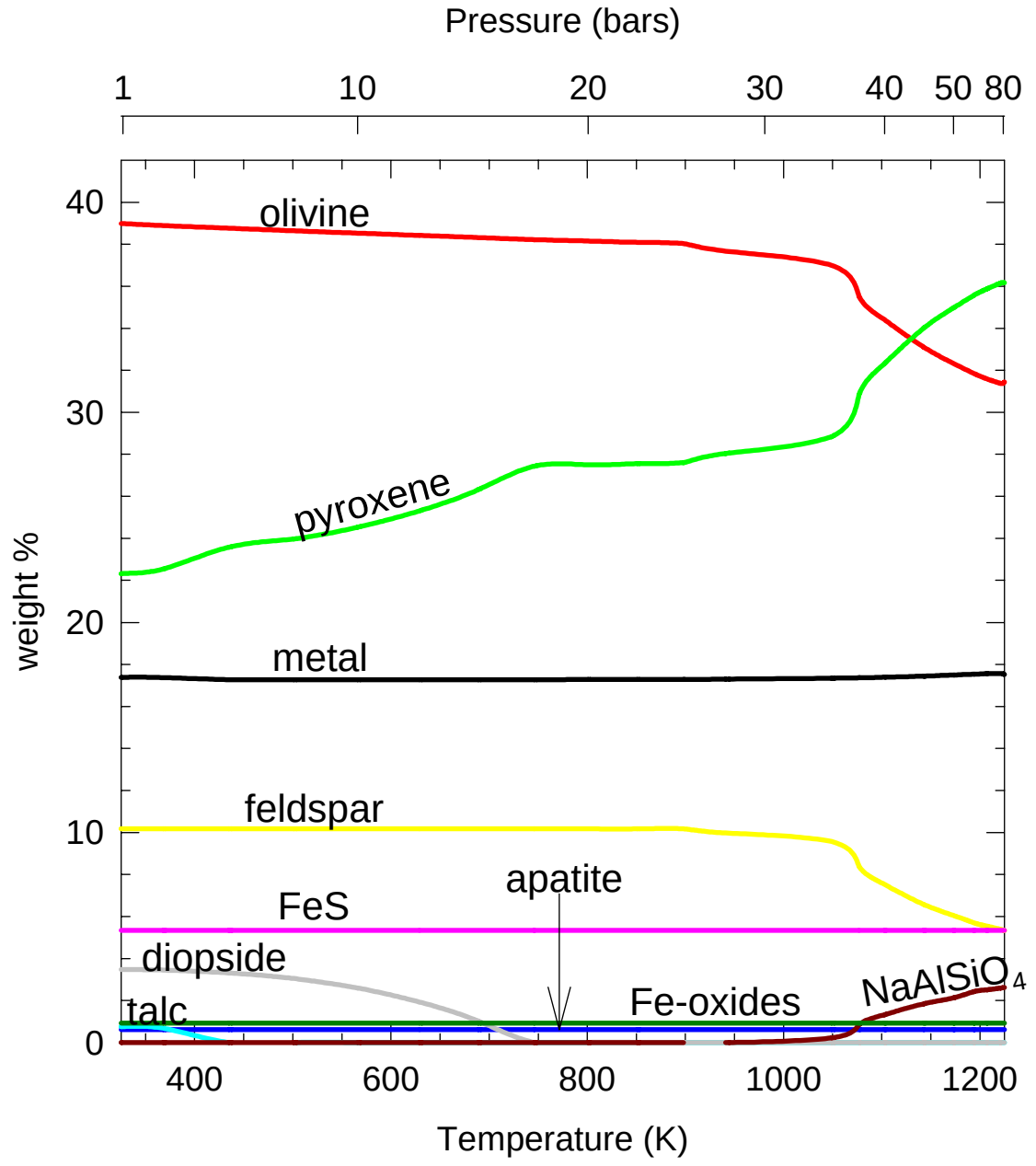

Figure 9. Schaefer and Fegley 2006

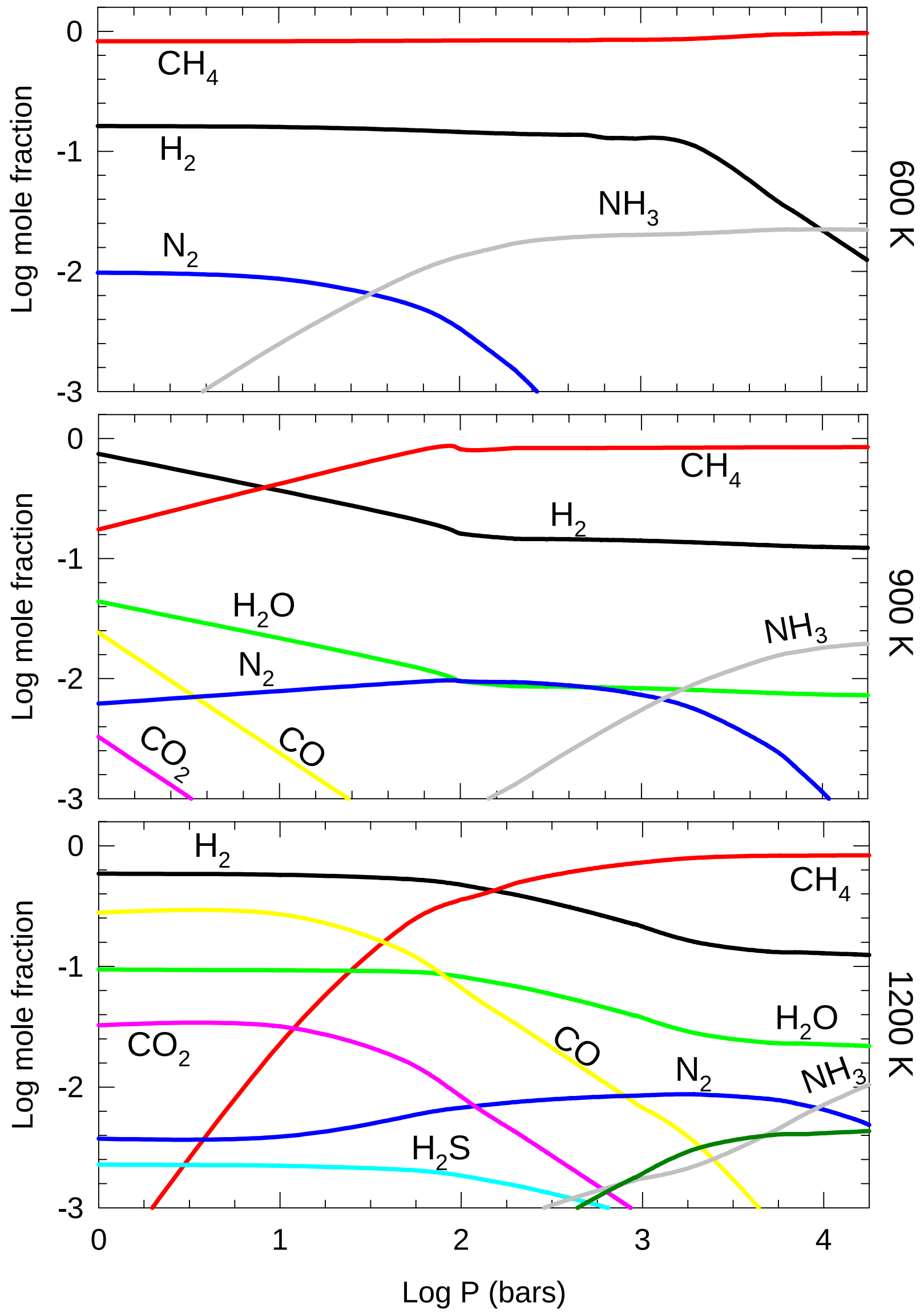

Figure 10. Schaefer and Fegley (2006)

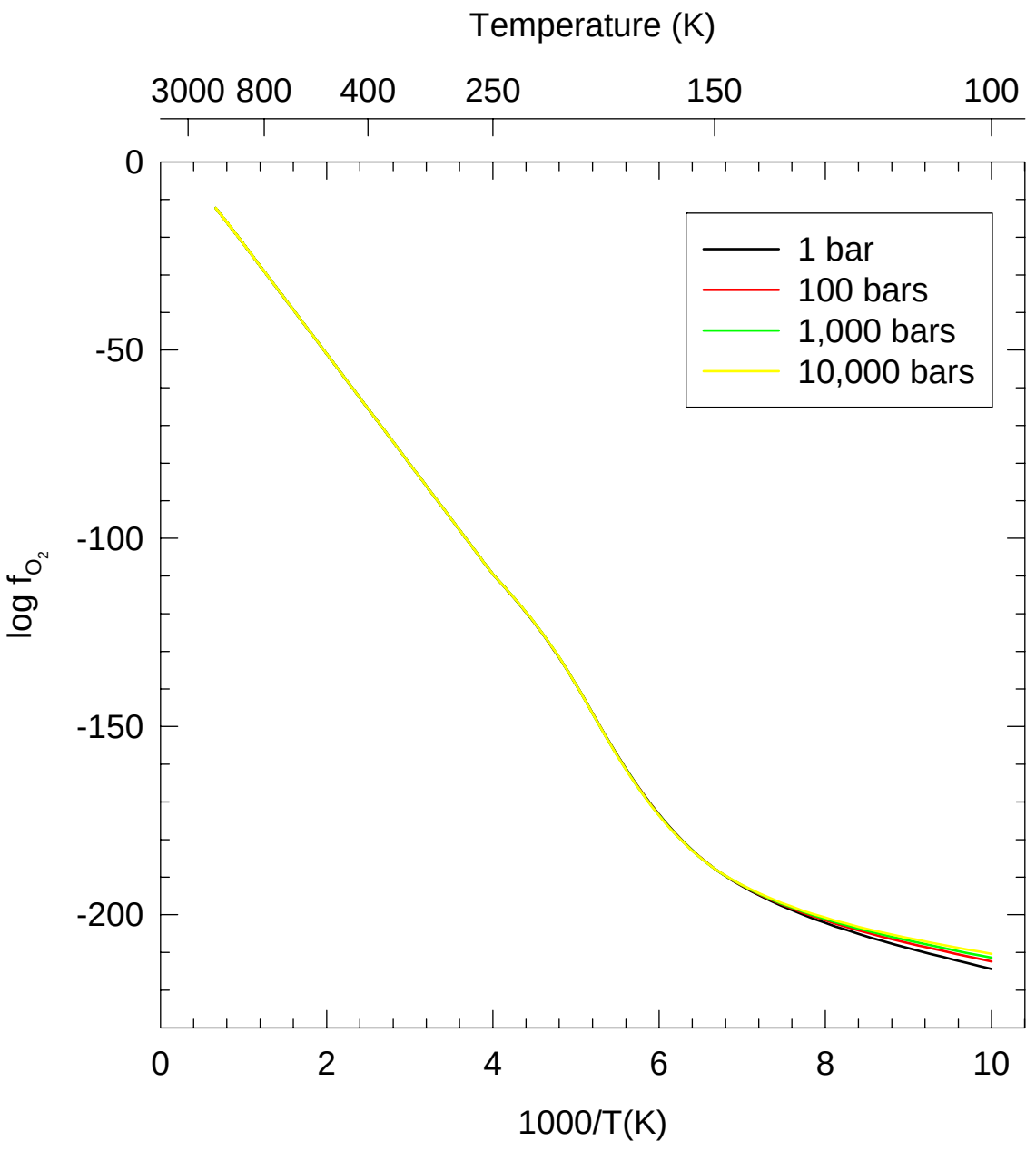

Figure 11. Schaefer and Fegley, 2006

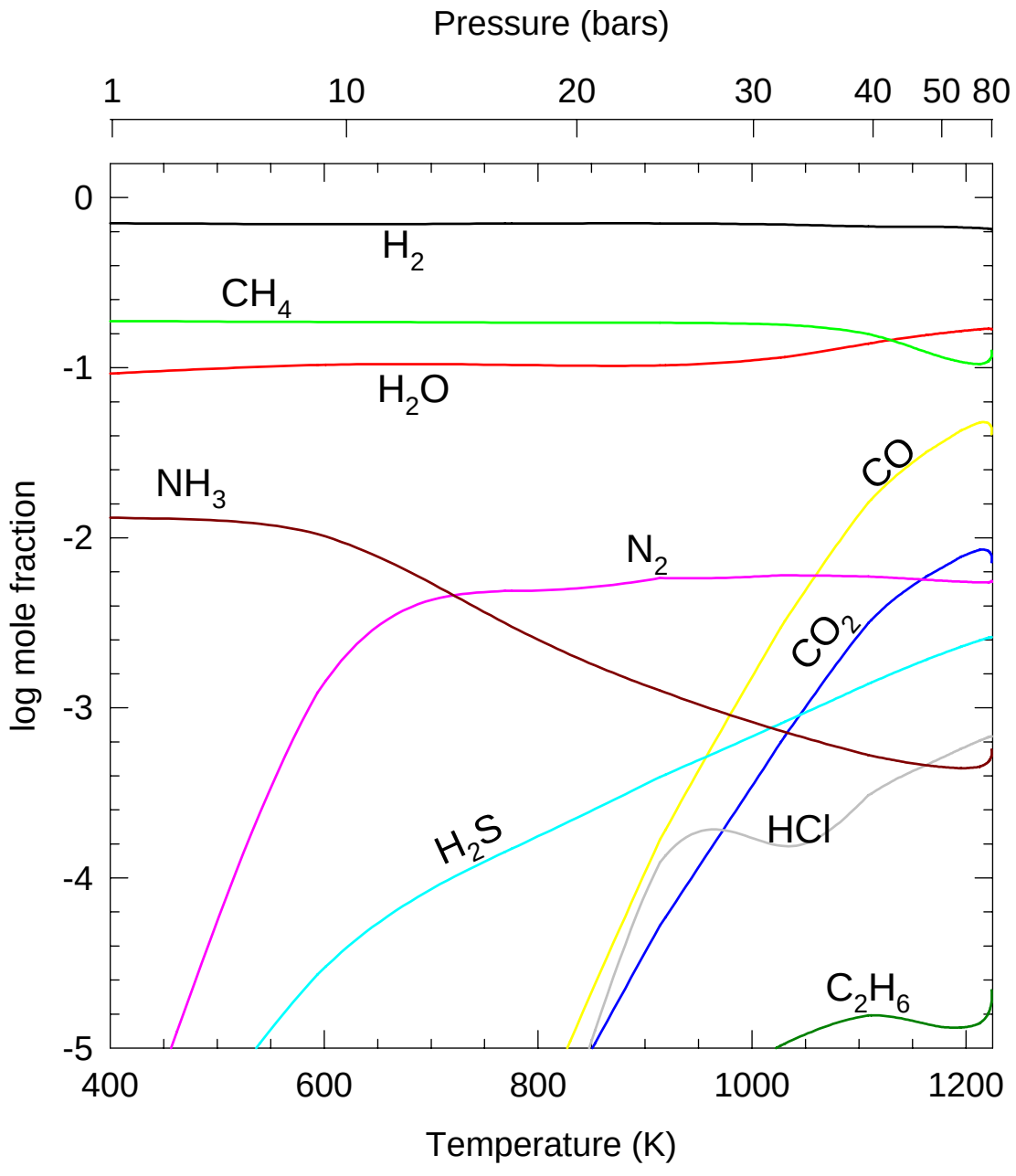

Figure 12. Schaefer and Fegley, 2006.

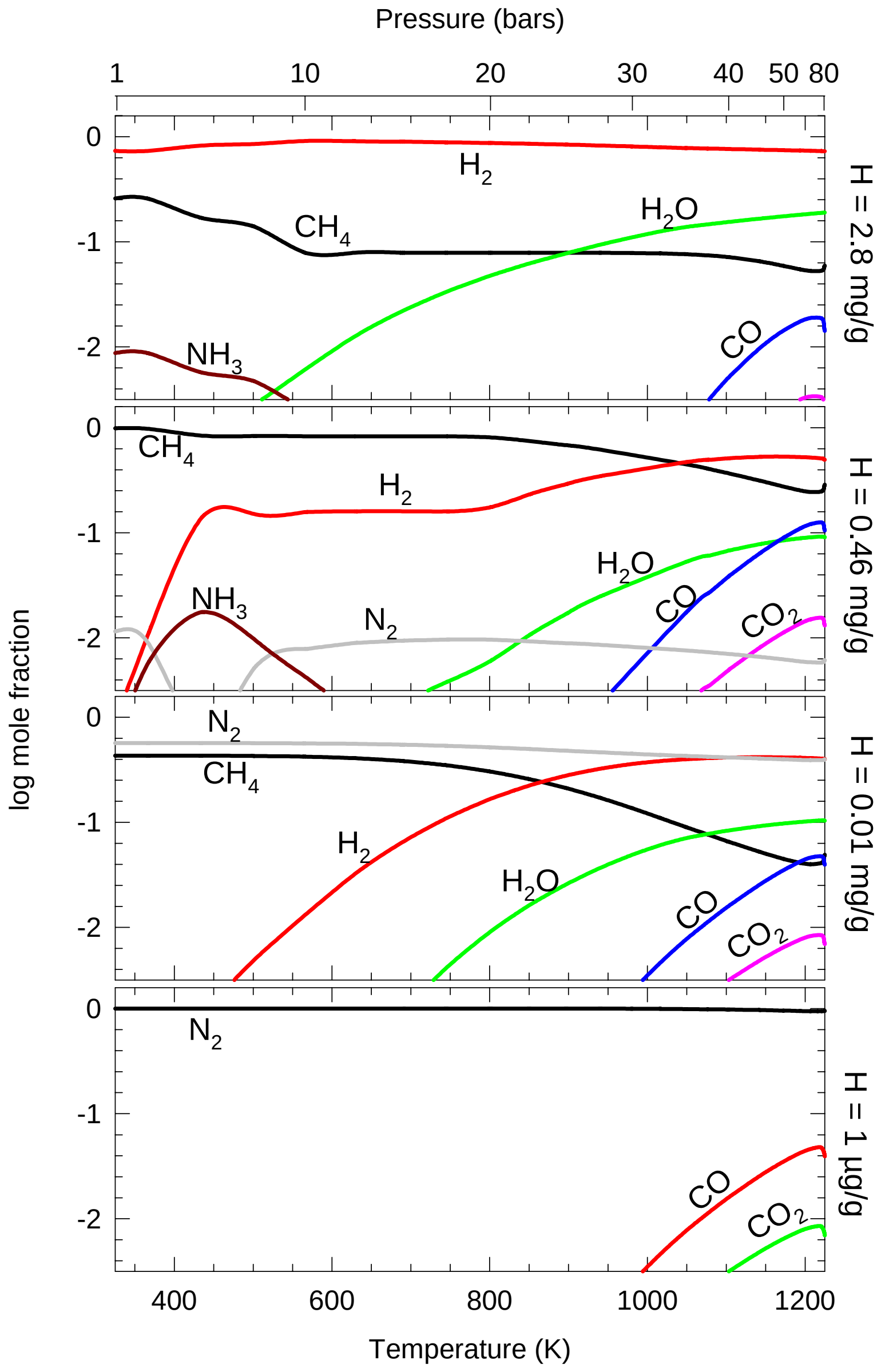

Figure 13. Schaefer and Fegley 2006

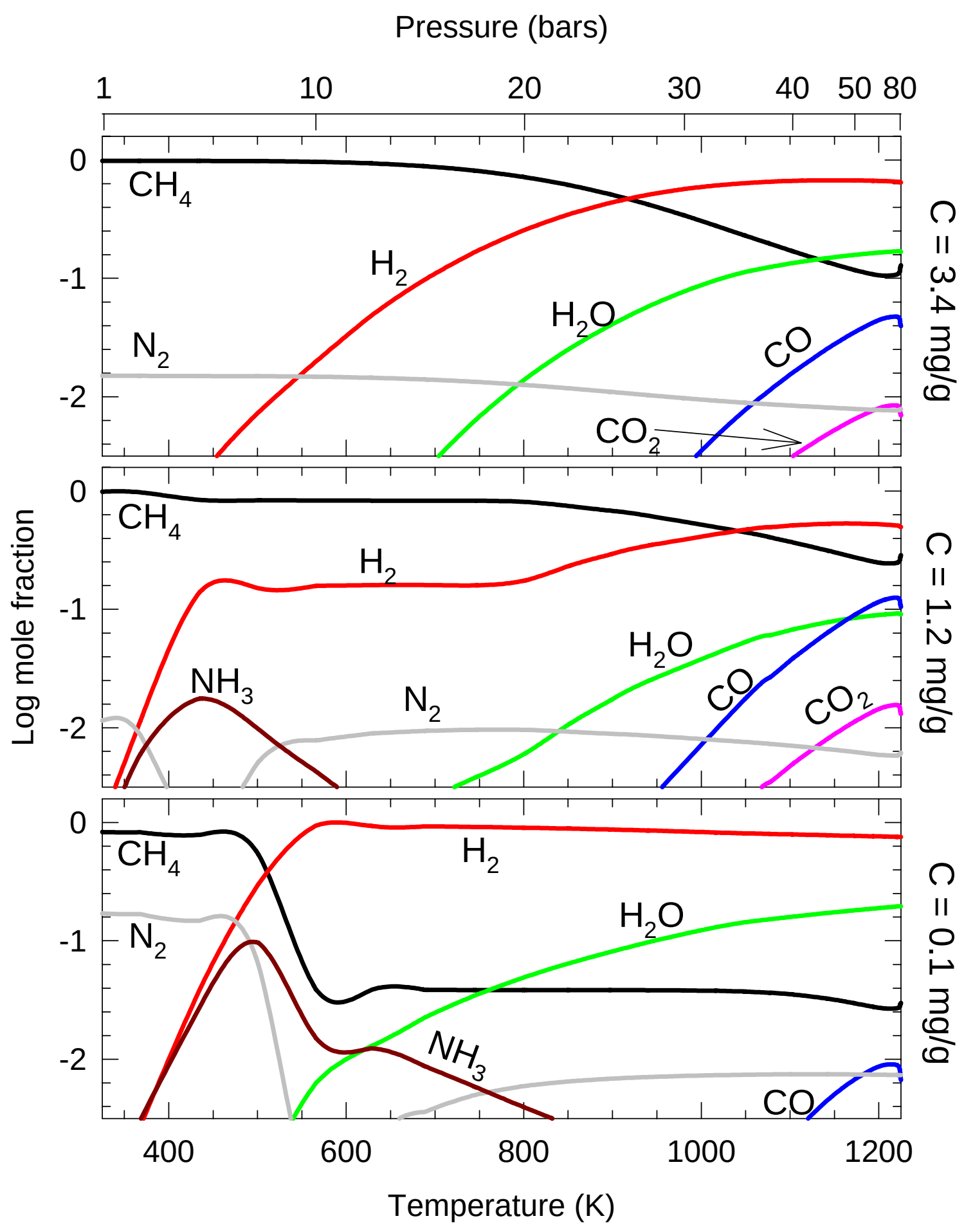

Figure 14. Schaefer and Fegley 2006

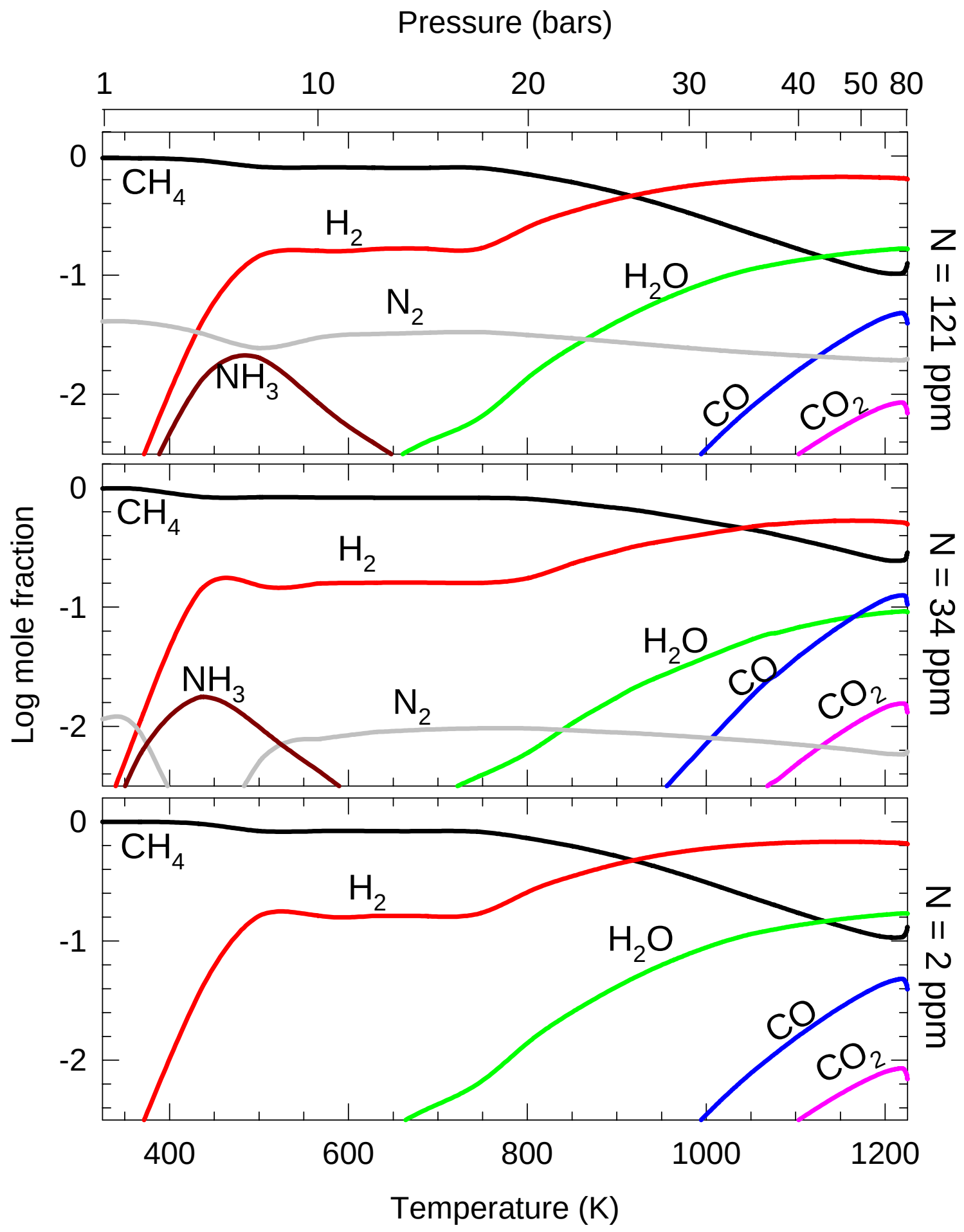

Figure 15. Schaefer and Fegley 2006

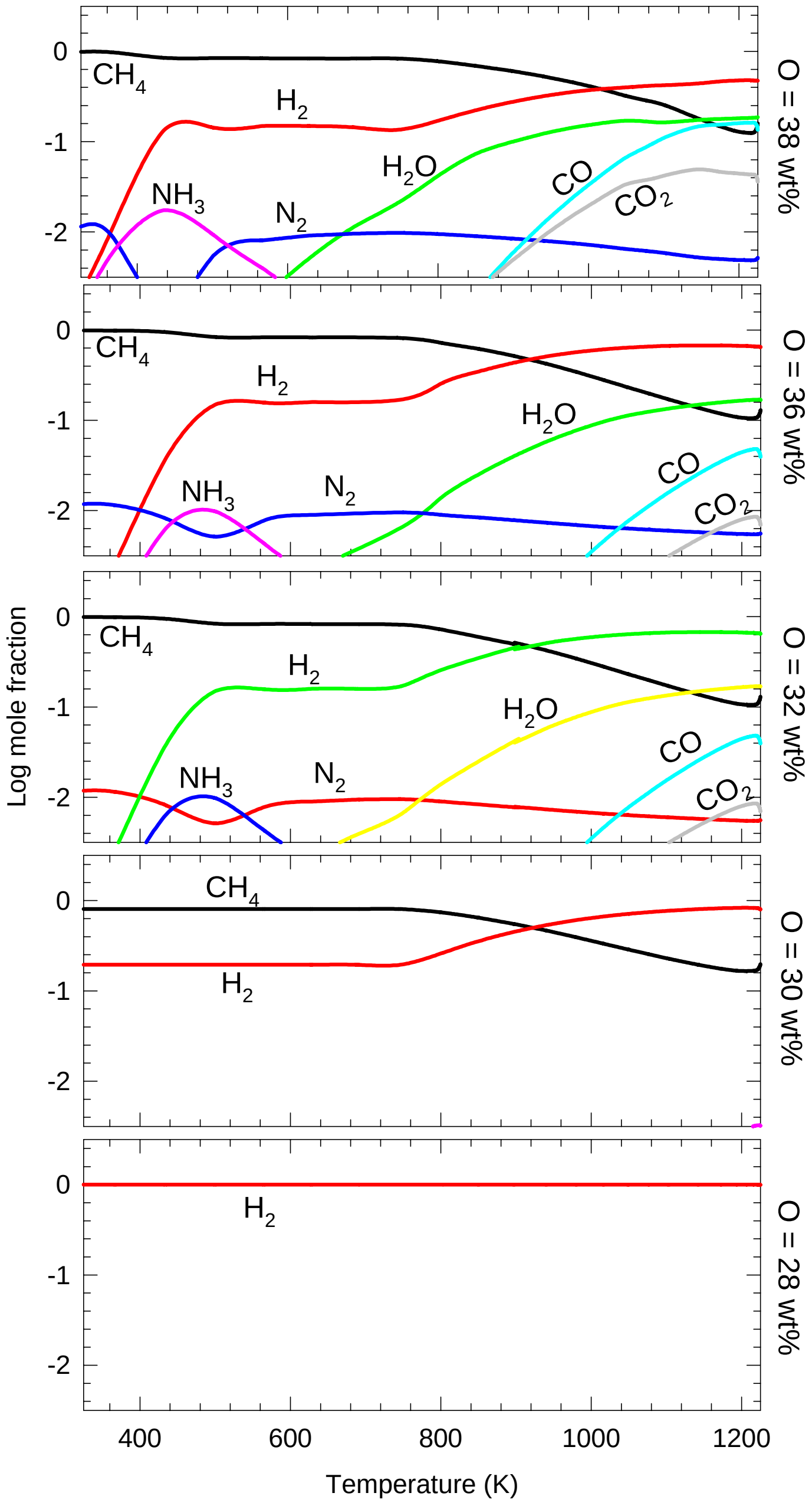

Figure 16. Schaefer and Fegley 2006.

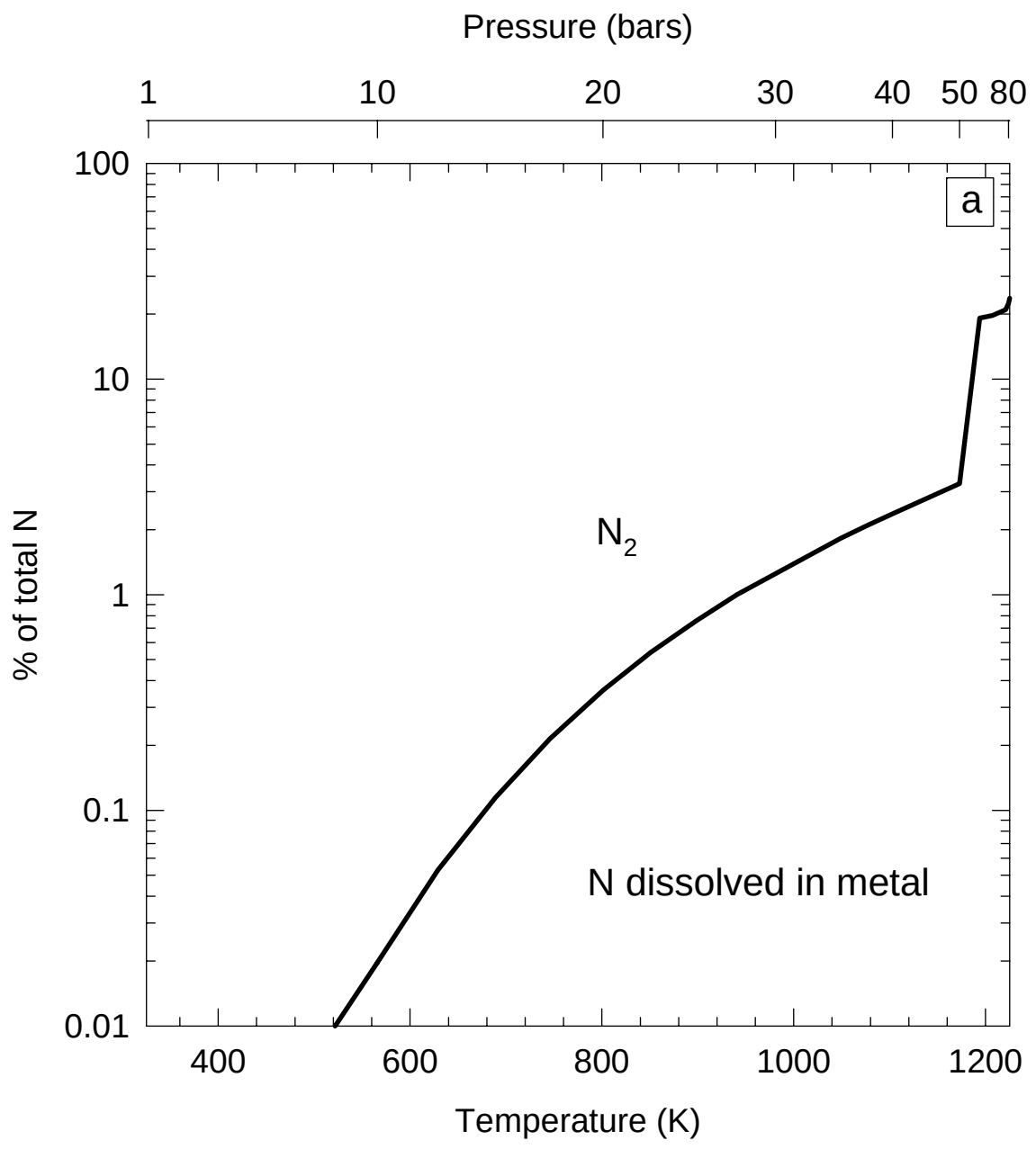

Figure 17(a) Schaefer and Fegley 2006

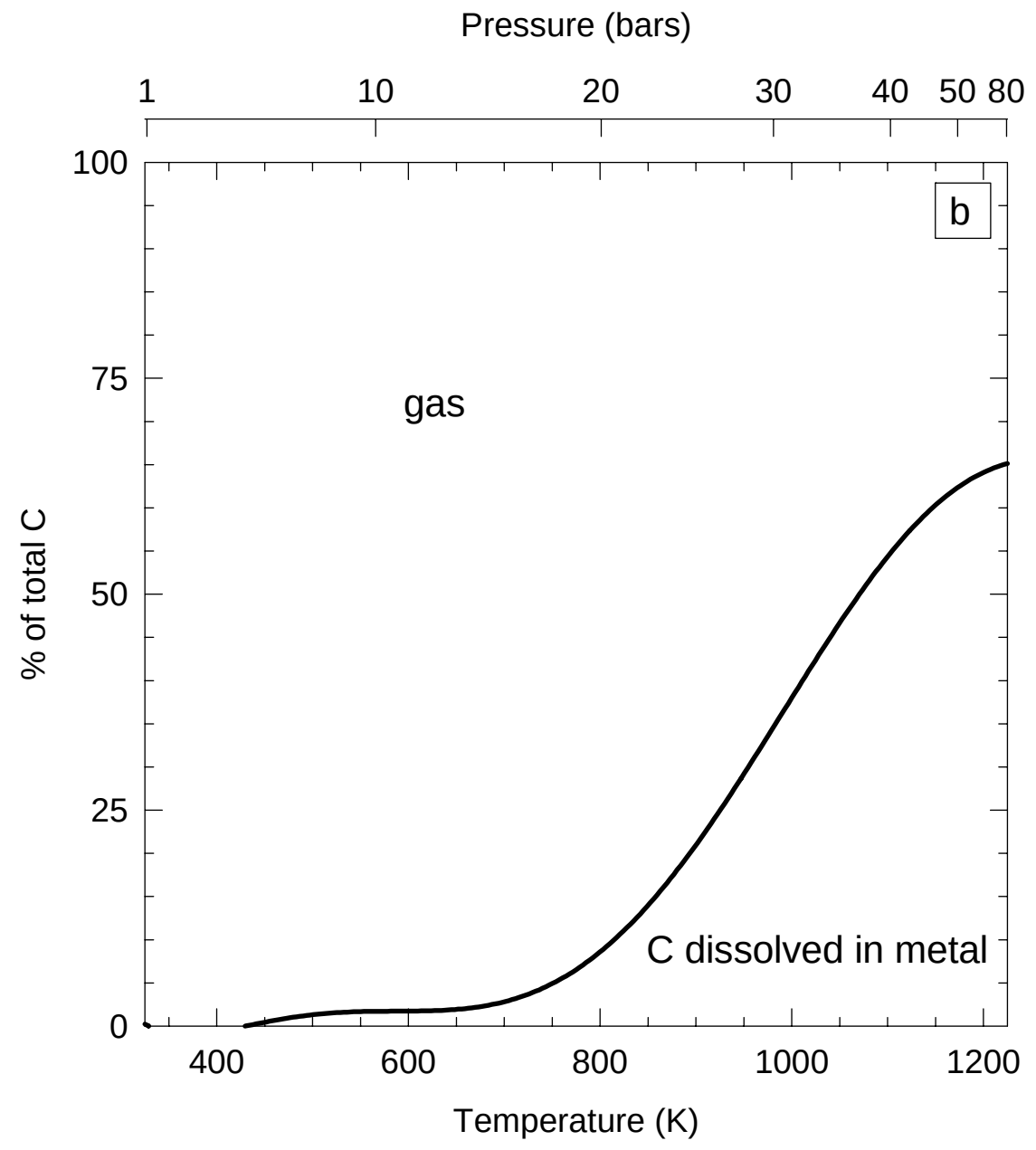

Fig. 17(b) Schaefer and Fegley 2006

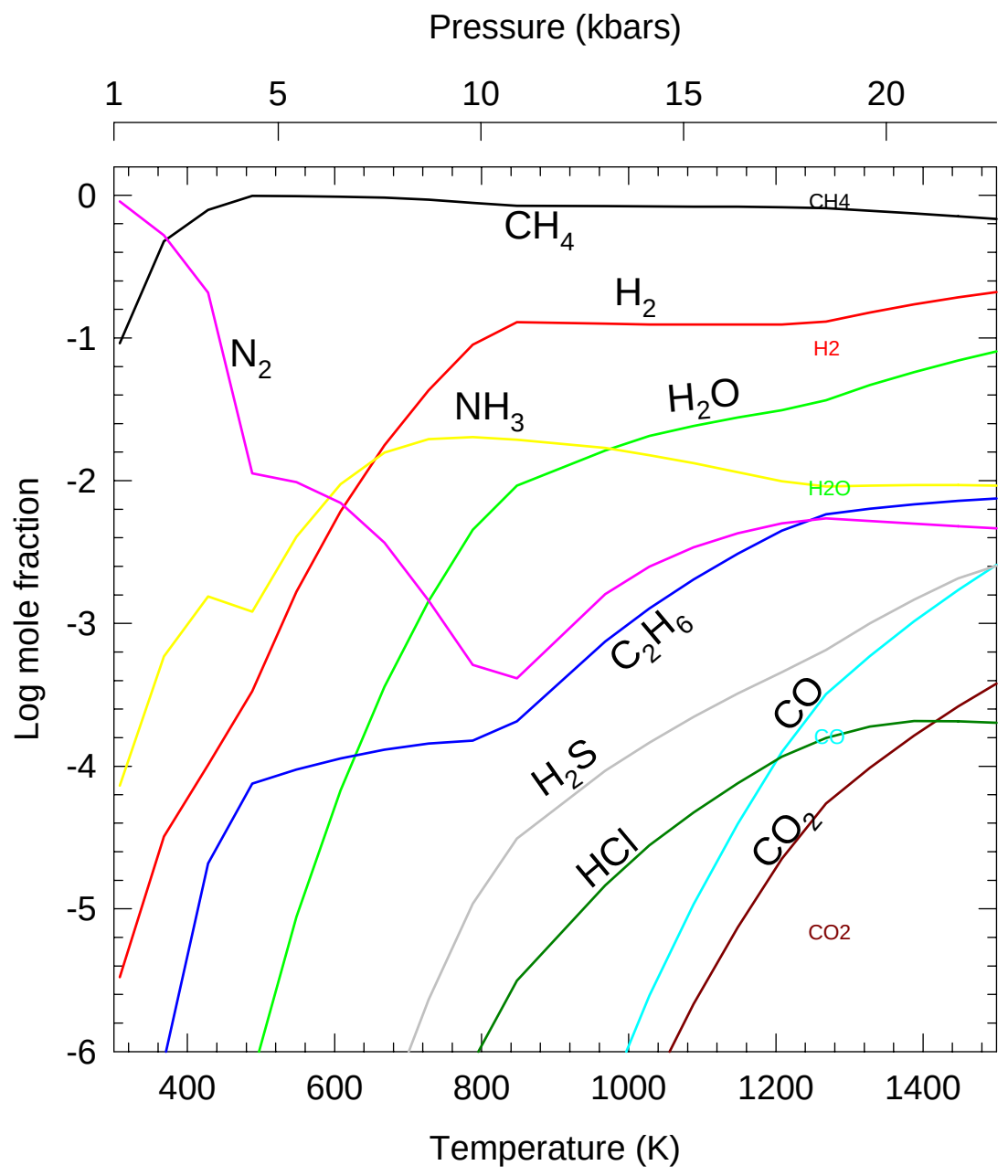

Figure 18, Schaefer and Fegley 2006